\documentclass[aps,prb,twocolumn,superscriptaddress,showpacs]{revtex4-1}

\usepackage{graphicx}
\usepackage{amsmath}
\usepackage{graphics}
\usepackage{exscale}
\usepackage{epsfig}

\renewcommand{\epsilon}{\varepsilon}

\newcommand{\integral}[3]{\!\int\limits_{#2}^{#3}\!\!{\rm d}#1\;}

\newcommand{\elcre}[2]{ c^{\dagger}_{#1,#2}}
\newcommand{\elann}[2]{ c_{#1,#2}}

\newcommand{\thGf}{{\cal G}}

\newcommand{\hc}{\mathrm{h.c.}}

\begin{document}

\title{Renormalized parameters and perturbation theory in dynamical mean-field theory for the Hubbard model}
\author{A.C. Hewson}
\affiliation{Department of Mathematics, Imperial College London, London SW7 2AZ,
  United Kingdom}
\date{\today} 
\begin{abstract}

We calculate the renormalized parameters for the quasiparticles and their interactions  for the Hubbard model in the paramagnetic phase as deduced from the low energy Fermi liquid fixed point using the results of a numerical renormalization group calculation (NRG) and dynamical mean-field theory (DMFT). Even in the low density limit
there is significant renormalization of the local quasiparticle interaction $\tilde U$, in agreement with
estimates based on the two-particle scattering theory of Kanamori (1963). On the approach to the Mott transition
we find a finite ratio for $\tilde U/\tilde D$, where $2\tilde D$ is the renormalized bandwidth, which is independent of whether the transition is approached by increasing the on-site interaction $U$ or on increasing the
density to half-filling.
The leading $\omega^2$ term in the self-energy and 
the local dynamical spin and charge susceptibilities are calculated  within the renormalized perturbation theory (RPT) and compared with the results calculated directly from the NRG-DMFT. We also suggest, more generally from the DMFT, how an
 approximate expression for the ${\bf q},\omega$ spin susceptibility  $\chi({\bf q},\omega)$ can derived from
repeated quasiparticle scattering with a local renormalized scattering vertex.

 \end{abstract}

\maketitle

\section{Introduction}
    
The  strong suppression of  charge fluctuations and enhancement of magnetic fluctuations 
in metallic systems with narrow energy bands, derived from atomic-like d or f states,
are a reflection of the strong renormalization of the low energy  quasiparticles in these systems. 
The extremely large effective
 masses, due to the very small quasiparticle weight factor $z$, has led to the classification of  many metallic rare earth and actinide metallic compounds as `heavy fermion'
systems\cite{Ste84,FSST95}. In some situations the quasiparticles disappear entirely at a quantum critical point
 as $z\to 0$ leading to finite temperature non-Fermi liquid behavior\cite{CPSR01,LRVW07,SS10}. In the cuprate superconductors
the apparent breakdown of Fermi liquid behavior appears to be closely  associated with a possible  electronic mechanism
for pairing leading to high temperature superconductivity in these materials\cite{Tai10,Pla10}. \par

 The basic mechanism 
driving these strong renormalization effects is believed in most cases to be the strong
local Coulomb interactions in the d or f shell orbitals. This renormalization is very well
understood in impurity systems where the strong local interaction is solely at the impurity
site, as described in the single impurity Anderson model. This understanding is based on
very effective non-perturbative techniques, such as the numerical renormalization group (NRG),
Bethe Ansatz (BA), conformal field theory (CFT), slave bosons and  $1/N$ expansions\cite{Wil75,TW83,AFL83,AL93,Col84,Hew93b}. 
The leading low energy effects can also be
calculated exactly in terms of quasiparticles and their interactions in a renormalized
perturbation theory\cite{Hew93,Hew06} (RPT). The breakdown of the quasiparticles
has also been described quantitatively in certain impurity models using these techniques\cite{NCH12a,NCH12b}.  
\par
 The corresponding generic lattice model describing electrons
in a narrow conduction  bands is the Hubbard model\cite{Hub63}.  Progress in understanding this model has
been much more limited, except for the model in one dimension, where an exact solution has been
obtained based on the Bethe Ansatz\cite{LW68,LW03}.  Models in one dimension, however, are known to be untypical of higher
dimensional systems as the low energy excitations are collective bose-like excitations, and
correspond to Luttinger liquids rather the Fermi liquids\cite{Hal81}. One non-perturbative technique, dynamical mean-field theory
(DMFT), has proved to be very effective in leading to an understanding of the metal to insulator, the Mott-Hubbard
transition, in the Hubbard and related models. 
 This approach is based on mapping the model  into an effective impurity model, which can then be solved using
an 'impurity solver'; the most commonly used being the numerical renormalization
group method\cite{BCP08} (NRG) or the Monte Carlo method\cite{Jar92,GK92} (MC). This mapping involves an approximation, but can be shown to 
be exact in the infinite dimensional limit, and to be a good approximation in systems where the self-energy
is strongly frequency dependent and has only a weak wavevector dependence, which is the usual situation in
three dimensional strongly correlated metals. 
 The earlier papers using this approach, with a detailed description of the application to the 
Mott-Hubbard transition  were reviewed in the article by Georges et al. \cite{GKKR96}. More recent developments have been the
application to models for particular metallic compounds, and  to include
finite dimensional effects which involve a mapping onto to an effective
cluster model rather than an impurity model\cite{PHK08,FGTJP11}.\par

Though there have been many studies of the Hubbard and related models using the
dynamical mean-field theory, the nature of the low energy quasiparticles
and their interactions has received little attention. In an earlier study we considered how
the quasiparticles for the Hubbard model vary in the presence of a magnetic field\cite{BH07b}
and also in an antiferromagnetic state \cite{BH07c}. There have been recent studies of the Hubbard\cite{LG16} and the related $t-J$ model\cite{SP16} concentrating 
 the region of the Mott-Hubbard transformation. It is of interest, therefore, to examine how the quasiparticles and their interactions are modified in this regime, as the quasiparticle weight $z\to 0$ on the approach
to the transition and the quasiparticles
disappear. Here we calculate
the quasiparticle renormalizations by analyzing the low energy NRG fixed point
from a DMFT-NRG calculation.
We can, not only characterize the free quasiparticles, but also deduce
the renormalized on-site quasiparticle interaction. The fact that the self-energy
of the effective impurity is the same as that for the on-site Green's function
of the lattice in the DMFT means   it can be calculated using the renormalized
perturbation theory for the effective impurity. This is one of the few analytic approaches
which is applicable in the strong correlation regime.  Some of the results, such as those
for the local spin and charge excitations, and the leading $\omega^2$ can be
checked against those deduced from the NRG calculations. However, expressions 
for ${\bf q}$ and $\omega$ dependent 
response functions, based on repeated quasiparticle scattering, go beyond the 
quantities that can be calculated directly using
the DMFT.\par

In  section \ref{s2} of the paper we give  background details of the model,
and the equations used in the DMFT and RPT. In section \ref{s3}  we survey the  results for the
renormalized parameters in the different regimes, and section \ref{s4}  look at the low
energy behaviour of the self-energy. In section \ref{s5}    we consider the
application of the RPT to the calculation of  local spin and charge
dynamic susceptibilities, and  in \ref{s6} suggest more generally how the corresponding ${\bf
  q}$ and $\omega$ dependent susceptibilities be might estimated from repeated quasiparticle scattering
with a local renormalized interaction vertex.
 Finally in section \ref{s7} we provide a summary and discuss
the possibilities for further developments using this approach.\par

\section{Dynamical mean-field approach and renormalized parameters\label{s2}}
The Hamiltonian for the single band  Hubbard model in a magnetic field  is given by
\begin{equation}
H_{\mu}=-\sum_{i,j,\sigma}(t_{ij}\elcre {i}{\sigma}\elann
{j}{\sigma}+\hc)-\sum_{i\sigma}\mu_{\sigma}
n_{i\sigma}+U\sum_in_{i,\uparrow}n_{i,\downarrow}\label{hubm}, 
\end{equation}
where $t_{ij}$ are the hopping matrix elements between sites $i$ and $j$,
$U$ is the on-site interaction; $\mu_{\sigma}=\mu+\sigma h$, where $\mu$ is the
chemical potential of the interacting system, and the Zeeman splitting term
with external magnetic field $H$ is given by $h=g\mu_{\rm B} H/2$, where
$\mu_{\rm B}$ is the Bohr magneton. 

From Dyson's equation, the one-electron Green's function $G_{{\bf k},\sigma}(\omega)$ can
be expressed in the form,  
\begin{equation}
G_{{\bf k},\sigma}(\omega) =\frac{1}{\omega+\mu_\sigma-\Sigma_{\sigma}({\bf
    k},\omega)-\epsilon({\bf k})},
\label{gk} 
\end{equation}
where $\Sigma_{\sigma}({\bf k},\omega)$ is the proper self-energy,
and $\epsilon({\bf k})=\sum_{\bf k}e^{-{\bf k}\cdot({\bf R}_i-{\bf R}_ j)}t_{ij}$.
The simplification that occurs for the model in the infinite dimensional limit
is that $\Sigma_{\sigma}({\bf k},\omega)$  becomes
a function of $\omega$ only \cite{MV89,Mue89}, so the local Green's function  
$ G_{\sigma}^{\mathrm{loc}}(\omega)$ takes the form,
\begin{equation}
G_{\sigma}^{\mathrm{loc}}(\omega) =\sum_{\bf k}G_{{\bf k},\sigma}(\omega)=
\integral{\epsilon}{}{}\frac{D(\epsilon)}
{\omega+\mu_\sigma -\Sigma_{\sigma}(\omega)-\epsilon},
\label{gloc}
\end{equation}
where $D(\epsilon)$ is the density of states for the non-interacting model
($U=0$).  In the dynamical mean-field theory approach \cite{GKKR96}, 
an auxiliary Green's function, $\thGf_{0,\sigma}(\omega)$, is introduced
 such that
\begin{equation}
  \thGf_{0,\sigma}^{-1}(\omega)=G_{\sigma}^{\mathrm{loc}}(\omega)^{-1}
  +\Sigma_{\sigma}(\omega),
\label{tgf}
\end{equation} 
 which can be written as
\begin{equation}
 G^{\mathrm{loc}}_{\sigma}(\omega) =
\frac{1}{ \thGf_{0,\sigma}^{-1}(\omega) -\Sigma_{\sigma}(\omega)}.
\label{locgf}
\end{equation}
This local Green's function   $ G_{\sigma}^{\mathrm{loc}}(\omega)$ can be  
identified as the Green's function 
$ G^{\rm imp}_{\sigma}(\omega)$ of an effective single impurity Anderson model,
and the
 auxiliary Green's function, $\thGf_{0,\sigma}(\omega)$, interpreted 
as the local Green's function for the {\em non-interacting} effective impurity.
If we
re-express $\thGf_{0,\sigma}^{-1}(\omega)$ in the form,
\begin{equation}
\thGf_{0,\sigma}^{-1}(\omega)=\omega+\mu+\sigma h-K_{\sigma}(\omega),
\label{thgfK}
\end{equation}
then Eqn. (\ref{locgf}) corresponds to the equation for
the impurity Green's function in a more conventional form,
\begin{equation}
G^{\rm imp}_{\sigma}(\omega)=\frac{1}
    {\omega-\epsilon_{\mathrm{d}\sigma}-K_\sigma(\omega)-\Sigma_\sigma(\omega)},
\label{gfdmft}
\end{equation}
 where  $\epsilon_{\mathrm{d}\sigma}=-\mu_\sigma$  plays the role of the
impurity level, and $K_\sigma(\omega)$ is the hybridization term.  In the impurity case
in the wide band limit  $K_{\sigma}(\omega)$ can be taken as $-i\Delta$
where $\Delta$ is a
constant. From
Eqns. (\ref{gloc}) and (\ref{tgf}) it follows that for the lattice model $K_\sigma(\omega)$ is a
function of the self-energy $\Sigma_\sigma(\omega)$. In the presence of an
applied  magnetic
field  it will also depend on the value of the field and on
$\sigma$.  As this self-energy is identified with the
impurity self-energy, which  in turn depends on  the form taken for
$K_\sigma(\omega)$, then  $K_\sigma(\omega)$ has to be determined
self-consistently and so plays the role of an effective
dynamical field. To define the model completely, we need to specify the  density of states $D(\omega)$ of the non-interacting
 model. For the infinite dimensional model this  is usually taken to be either 
that for a
tight-binding hypercubic or that for a Bethe lattice. Here we take the
 semi-elliptical  form corresponding to a Bethe lattice,
\begin{equation}
    D(\epsilon,\mu)=\frac{2}{\pi D^2}\sqrt{D^2-(\epsilon+ \mu)^2},
\label{dos}
 \end{equation}
where $2D$ is the band width, with $D=2t$ for the Hubbard model, and $\mu$
the chemical potential of the free electrons. We choose
this form with the value $t=1$ throughout, rather than the Gaussian density of states of the hypercubic lattice,
as it has a finite bandwidth ($W=4t=4.0$).\par 
The focus here will be on using the renormalized perturbation theory (RPT) in the strongly correlated regime where standard perturbation theory
is not applicable.

 In formulating  RPT approach we assume that the self-energy
  $\Sigma_\sigma(\omega)$ can be written in the form 
\begin{equation}
 \Sigma_\sigma(\omega)= \Sigma_\sigma(0)+\omega \Sigma'_\sigma(0)
 +\Sigma_\sigma^{\rm rem}(\omega),
\label{selfexp}
\end{equation}
which corresponds to an expansion in powers of $\omega$ to first order but includes a remainder
term $\Sigma_\sigma^{\rm rem}(\omega)$. We assume the Luttinger
result that the imaginary part of the self-energy behaves asymptotically
as $\omega^2$ as $\omega\to 0$, so that both $\Sigma_\sigma(0)$
and $\Sigma'_\sigma(0)$ can be taken to be real\cite{Lut60}. These two 
assumptions imply that the low energy fixed point corresponds 
to a Fermi liquid. No terms have been omitted so, apart from these assumptions,
there is no approximation involved. Substituting this form for the self-energy
into Eqn. (\ref{gk}), it can be written in the form
\begin{equation}
G_{{\bf k},\sigma}(\omega) =\frac{z_\sigma}{\omega+\tilde\mu_\sigma-\tilde\Sigma_{\sigma}(\omega)-\tilde\epsilon_\sigma({\bf k})},
\label{rgk} 
\end{equation}
where
\begin{equation} \tilde\mu_{\sigma}=z_{\sigma}(\mu_{\sigma}-\Sigma_{\sigma}(0)),\quad
     z_{\sigma}=1/[1-\Sigma'_{\sigma}(0)],
\label{nrgqp}
\end{equation}
 $\tilde\epsilon(\sigma,{\bf k})=z_\sigma\epsilon({\bf k})$
and $\tilde\Sigma_{\sigma}(\omega)$ is the renormalized self-energy
defined by 
 \begin{equation}
\tilde\Sigma_{\sigma}(\omega)=z_\sigma\Sigma^{\rm rem}_{\sigma}(\omega).
\end{equation}
We interpret $z_\sigma$ as a quasiparticle weight factor, and define
a quasiparticle Green's function, $\tilde G_{{\bf k},\sigma}(\omega)$,
for the interacting system as
\begin{equation}
\tilde G_{{\bf k},\sigma}(\omega) =\frac{1}{\omega+\tilde\mu_\sigma-\tilde\Sigma_{\sigma}(\omega)-\tilde\epsilon_\sigma({\bf k})},
\label{gkqp} 
\end{equation}
which is now similar in form to that given in Eqn. (\ref{gk}).
The free quasiparticle Green's function, $\tilde G_{{\bf k},\sigma}(\omega)$,
 corresponds to putting $\tilde\Sigma_{\sigma}(\omega)=0$ in Eqn.
 (\ref{gkqp}).
\par
Using the same expression for the self-energy in the local Green's function
(\ref{gloc}), it can be rewritten in the form,
\begin{equation}
 G_{\sigma}^{\mathrm{loc}}(\omega) =z_\sigma\integral{\epsilon}{}{}\frac{D(\epsilon/z_{\sigma})}
{\omega+\tilde\mu_{\sigma} -\epsilon-\tilde\Sigma_\sigma(\omega)}.
\label{gqploc}
\end{equation}
The local 
free quasiparticle propagator, 
 $G_{0,\sigma}^{\mathrm{loc}}(\omega)$, is given by
\begin{equation}
\tilde G_{0,\sigma}^{\mathrm{loc}}(\omega) =\integral{\epsilon}{}{}\frac{D(\epsilon/z_{\sigma})}
{\omega+\tilde\mu_{\sigma} -\epsilon}.
\label{fgqploc}
\end{equation}
The  density of states  $\tilde \rho_{\sigma}(\omega)$  derived from this Green's function
 via $\tilde \rho_{\sigma}(\omega)=-{\rm Im}\tilde
G_{0,\sigma}(\omega+i\delta)/\pi$ we will refer to as the free quasiparticle
density of states (DOS).
 For the Bethe lattice, this DOS
takes the form of a band with renormalized parameters,
\begin{equation}
   \tilde \rho_{\sigma}(\omega)=\frac{2}{\pi\tilde D_{\sigma}^2}\sqrt{\tilde 
   D_{\sigma}^2-(\omega+\tilde \mu_{\sigma})^2},
\label{qpdos}
 \end{equation}
where $\tilde D_{\sigma}=z_{\sigma}D$. \par

The renormalized perturbation theory is set up such that the
propagators used in the expansion correspond to the fully dressed
non-interacting quasiparticles, and the expansion is in powers of
the quasiparticle interaction which is identified with full four-vertex
between spin up and spin down electrons on the same site $i$
 evaluated with all the frequency arguments set to zero,
\begin{equation}\tilde U=z_\uparrow z_\downarrow\Gamma^{(4)}_{i\uparrow,i\downarrow,i\downarrow,i\uparrow}(0,0,0,0).\end{equation}
This vertex with zero frequency arguments is well defined in the finite frequency $T=0$ perturbation theory,
and being a local vertex with the all site indices corresponding to a single site is the same for
the effective impurity and lattice in the infinite dimensional limit. 
  Counter terms
must be included in the calculation  to cancel off any 
 renormalizations which may be generated in the
expansion. As the quasiparticles are taken to be fully renormalized
any further renormalization would result in
overcounting.\par
We will need 
the values of the renormalized parameters to substitute in the RPT  and these we deduce from the NRG  calculation
for the effective impurity.
We first consider how to calculate the parameters $z_\sigma$ and
$\tilde\mu_{\sigma}$ which characterise the free quasiparticles.
For the NRG calculations for the Anderson model the conduction electron density of
states is discretized and transformed into a form which corresponds to  a one dimensional tight
binding chain. This conduction electron chain is then  coupled via an effective hybridization $V_\sigma$ to the impurity \cite{KWW80a}. 
 In this representation $K_\sigma(\omega)=|V_\sigma|^2
g_{0,\sigma}(\omega)$, where $g_{0,\sigma}(\omega)$ is the one-electron Green's
function for the first site of the isolated conduction electron chain.
We substitute the self-energy
$\Sigma_\sigma(\omega)$ into the form given earlier into Eqns. (\ref{selfexp}) 
  and (\ref{gfdmft}),
\begin{equation}
G^{\rm imp}_{\sigma}(\omega)=\frac{z_\sigma}{\omega-\tilde\epsilon_{\mathrm{d}\sigma}-|\tilde
  V_\sigma|^2g_{0,\sigma}(\omega)-\tilde\Sigma_\sigma(\omega)}, 
\label{fqpgfdmft}
\end{equation}
where
\begin{equation}
\tilde\epsilon_{\mathrm{d}\sigma}=z_\sigma(\epsilon_{\mathrm{d}\sigma}+\Sigma_\sigma(0)),\quad
|\tilde V_\sigma|^2={z_\sigma}| V_\sigma|^2.
\label{rp}
\end{equation}
The corresponding free quasiparticle impurity Green's function, $\tilde G^{\rm
  imp}_{0,\sigma}(\omega)$,
is then given by
\begin{equation}
\tilde G^{\rm imp}_{0,\sigma}(\omega)=\frac{1}{\omega-\tilde\epsilon_{\mathrm{d}\sigma}-|\tilde
  V_\sigma|^2g_{0,\sigma}(\omega)}.
\label{qpgfdmft}
\end{equation}
As we identify $ G^{\rm imp}_{\sigma}(\omega)$ with the local  Green's
function for the lattice (\ref{gloc}), it follows that
\begin{equation}
  \tilde G_{0,\sigma}^{\mathrm{loc}}(\omega)=\tilde G^{\rm imp}_{0,\sigma}(\omega),
\end{equation}
which  specifies the form of $g_{0,\sigma}(\omega)$ in (\ref{qpgfdmft}) 
and yields $\tilde\mu_{\sigma}=-\tilde\epsilon_{\mathrm{d}\sigma}$.
By fitting the lowest lying poles of this Green's function to the lowest lying
single particle and hole excitations in the NRG results,
 we can deduce  the parameters $\tilde\epsilon_{\mathrm{d}\sigma}$
and $\tilde V_\sigma$, as has been explained in earlier work. 
\cite{HOM04}. The quasiparticle weight $z_\sigma$ is then obtained from the
relation $z_\sigma=|\tilde V_\sigma/ V_\sigma|^2$ in Eqn. (\ref{rp}), and $\tilde\mu_{\sigma}$ from
$\tilde\mu_{\sigma}=-\tilde\epsilon_{\mathrm{d}\sigma}$.\par
We also need to calculate the renormalized on-site interaction $\tilde U$ for
the effective impurity. This can be deduced from the difference in energies 
between the lowest lying two-particle excitation from the NRG ground state and the corresponding 
two free single particle excitations. This procedure is
difficult to summarize, so we refer to the earlier work for details in Ref.~\onlinecite{HOM04}.   

\section{Results for Renormalized Parameters \label{s3}}
\begin{figure}[!htbp]
\begin{center}
\includegraphics[width=0.45\textwidth]{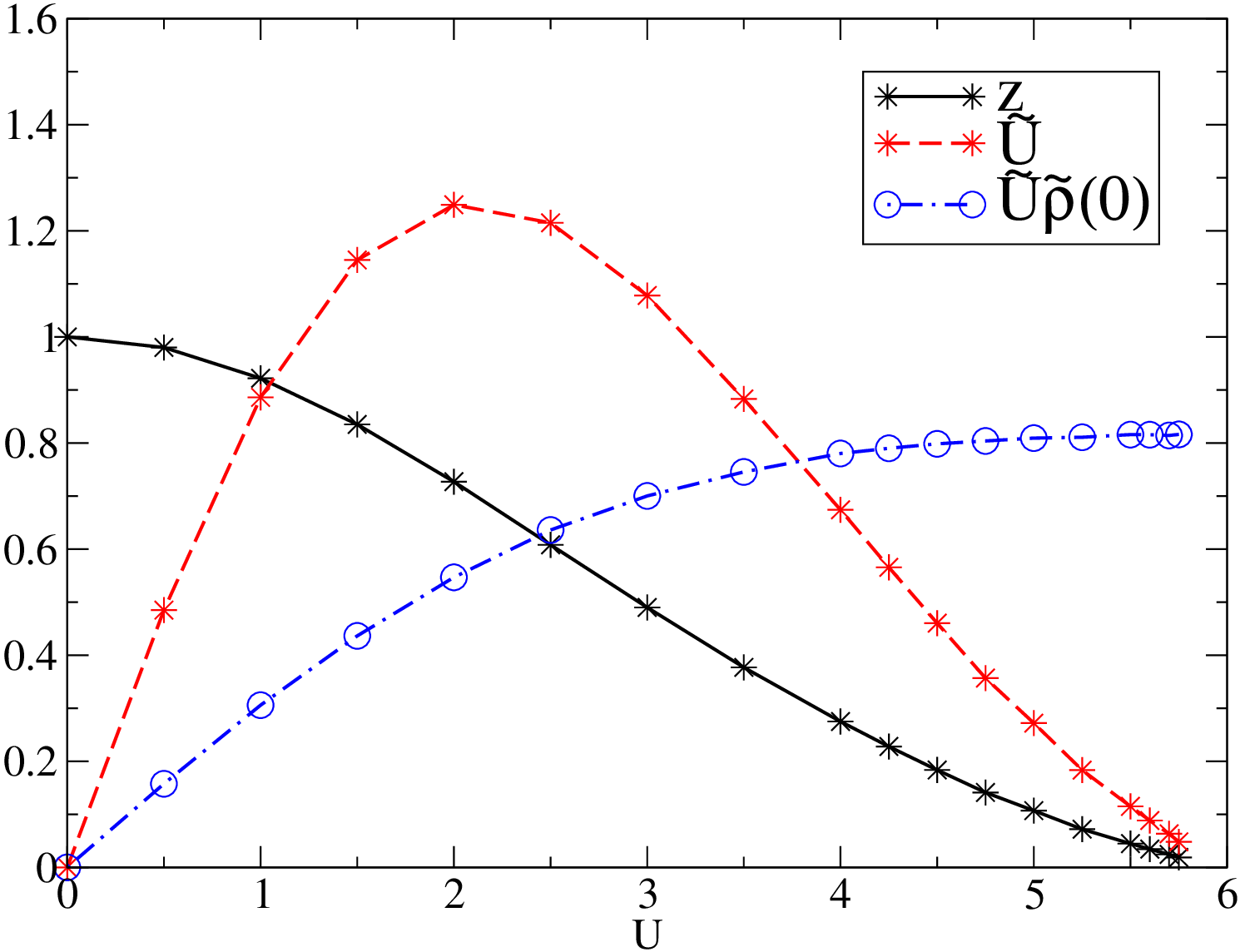}

\caption{ The quasiparticle weight
  $z=\tilde D/D$, the on-site quasiparticle interaction $\tilde U$, and the
  product $\tilde U\tilde\rho(0)$ for the
  model at half-filling as a function of 
$U$  }
\label{rp1}
 \vspace*{0.5cm}
\end{center}
\end{figure}

Here we use the NRG method to solve the DMFT equations for the effective
impurity to calculate the renormalized parameters $z=\tilde D/D$, $\tilde
\mu_{\sigma}$ and $\tilde U$ in different parameter regimes. In the
half-filled
case in the absence of a magnetic field $\tilde \mu_{\sigma}=0$, so  we have
just two parameters to determine, $z=\tilde D/D$ and $\tilde U$. These are plotted
as a function of $U$ in Fig. \ref{rp1}. For small $U$, $\tilde U$
is, as expected, proportional to $U$ up to a value of $U\sim 1$.
As the Mott transition is approached at a critical
value $U_c=5.98$\cite{Bul99} (as $D=2$ in our case $U_c/D=2.99$), it can be seen that both $\tilde U$ and $z$ approach zero
in a similar way. If we form the dimensionless ratio $\tilde
U\tilde\rho(0)$,
then with $\tilde\mu=0$, $\tilde\rho(0)=2/\pi\tilde D$, we find that $\tilde
U\tilde\rho(0)\to 0.815$ as $U\to U_c$. 
We also see from Eqn. (\ref{qpdos}) that as $\tilde D\to 0$, that the quasiparticle density of states
narrows to a delta function at $\omega=0$ as $U\to U_c$.\par

We can define a quasiparticle occupation number  $ \tilde
 n_{\sigma}$  at $T=0$,  by integrating the free quasiparticle density of states up to the Fermi
 level
\begin{equation}
   \tilde n_{\sigma}=\int_{-\infty}^{0} d\omega \tilde\rho_{\sigma}(\omega).
\label{qpocc}
 \end{equation}

\noindent

We can also calculate  the expectation value of the 
occupation number $n_{\sigma}$  of the interacting system at $T=0$ using a
generalization of Luttinger's theorem \cite{LW60} for each spin component,
 \begin{equation}
   n_{\sigma}=\integral{\epsilon}{-\infty}{\infty} D(\epsilon)
\theta(\mu_\sigma-\Sigma_\sigma(0)-\epsilon),
\label{nocc}
 \end{equation}
where $\theta(\epsilon)$ is the Heaviside step function and $D(\epsilon)$ as given
in Eqn. (\ref{dos}). It can be shown that this result is equivalent to
that given in Eqn. (\ref{qpocc}) 
so  $\tilde n_{\sigma}= n_{\sigma}$, and hence we can calculate the occupation
number $  n_{\sigma}$ from the quasiparticle density of states  $\tilde\rho_{\sigma}(\omega)$.\par
We can evaluate the integral in Eqn. (\ref{qpocc}) explicitly in the 
case of a semi-elliptical density of states, which gives
\begin{equation}
   \tilde n_{\sigma}=n_{\sigma}={1\over\pi}\left[{\pi\over 2}+{\rm
   sin}^{-1}\left({\tilde\mu_{\sigma}\over\tilde D}\right)+{\tilde \mu_{\sigma}\over \tilde
   D^2}\sqrt{\tilde D^2-\tilde\mu_{\sigma}^2}\right].
\label{noccex}
 \end{equation}
The magnetization $m(h)$ can be deduced from (\ref{noccex}) using
$m(h)=g\mu_{\rm B}(n_{\uparrow}-n_{\downarrow})$.
In the half-filled case and in the absence of a magnetic field, $\tilde
\mu=0$, and we see that $  \tilde n_{\sigma}
=0.5$, so it is even possible to assign a value in the localized limit
when $z\to 0$, and the quasiparticle density of states collapses to a delta-function.

\begin{figure}[!htbp]
\begin{center}
\includegraphics[width=0.45\textwidth]{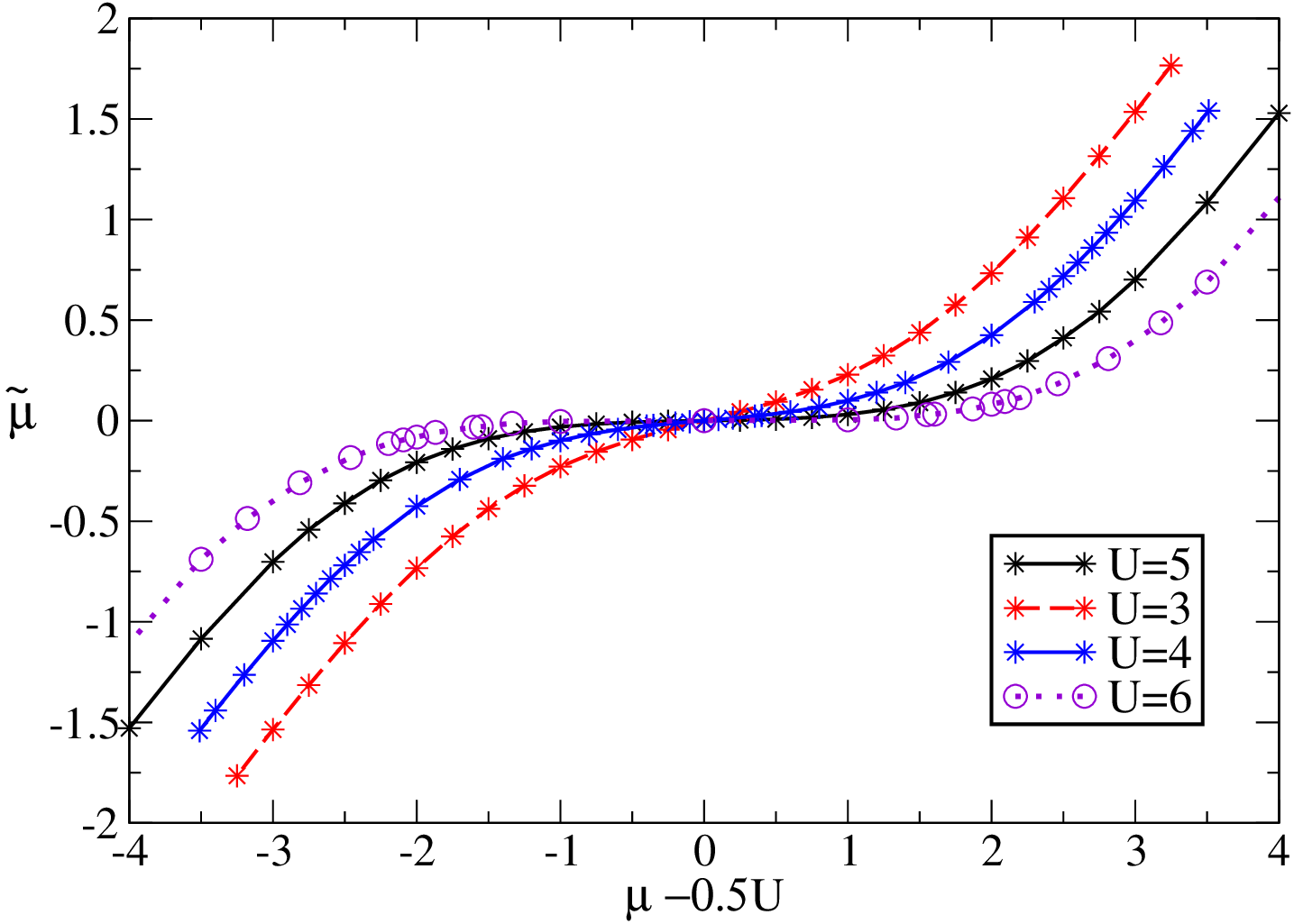}
\caption{The quasiparticle chemical potential $\tilde \mu$ as a function of
  the on-site occupation $n$ plotted as a function of $\bar\mu=\mu-U/2$ for
$U=3,4,5,6$.  }
\label{tildemu_vs_mu}
\vspace*{0.5cm}
\end{center}
\end{figure}

In Fig. \ref{tildemu_vs_mu} we give a plot of the renormalized
chemical potential $\tilde \mu$ as a function of $\bar \mu=\mu-0.5U$
($\bar\mu=0$ corresponds to the particle-hole symmetry) for
$U=3, 4,5,6$. For these values up to $U=5$ we are in the metallic regime,
so $\tilde\mu$ is a continuous function of $\bar\mu$, but as $U$ increases 
there a plateau region develops about $\bar\mu=0$, corresponding to a strong
correlation regime and a reduced charge susceptibility. For $U=6$ we are very slightly
above the critical value $U=5.98$ but so close that the discontinuity is not evident.

\begin{figure}[!htbp]
\begin{center}
\includegraphics[width=0.45\textwidth]{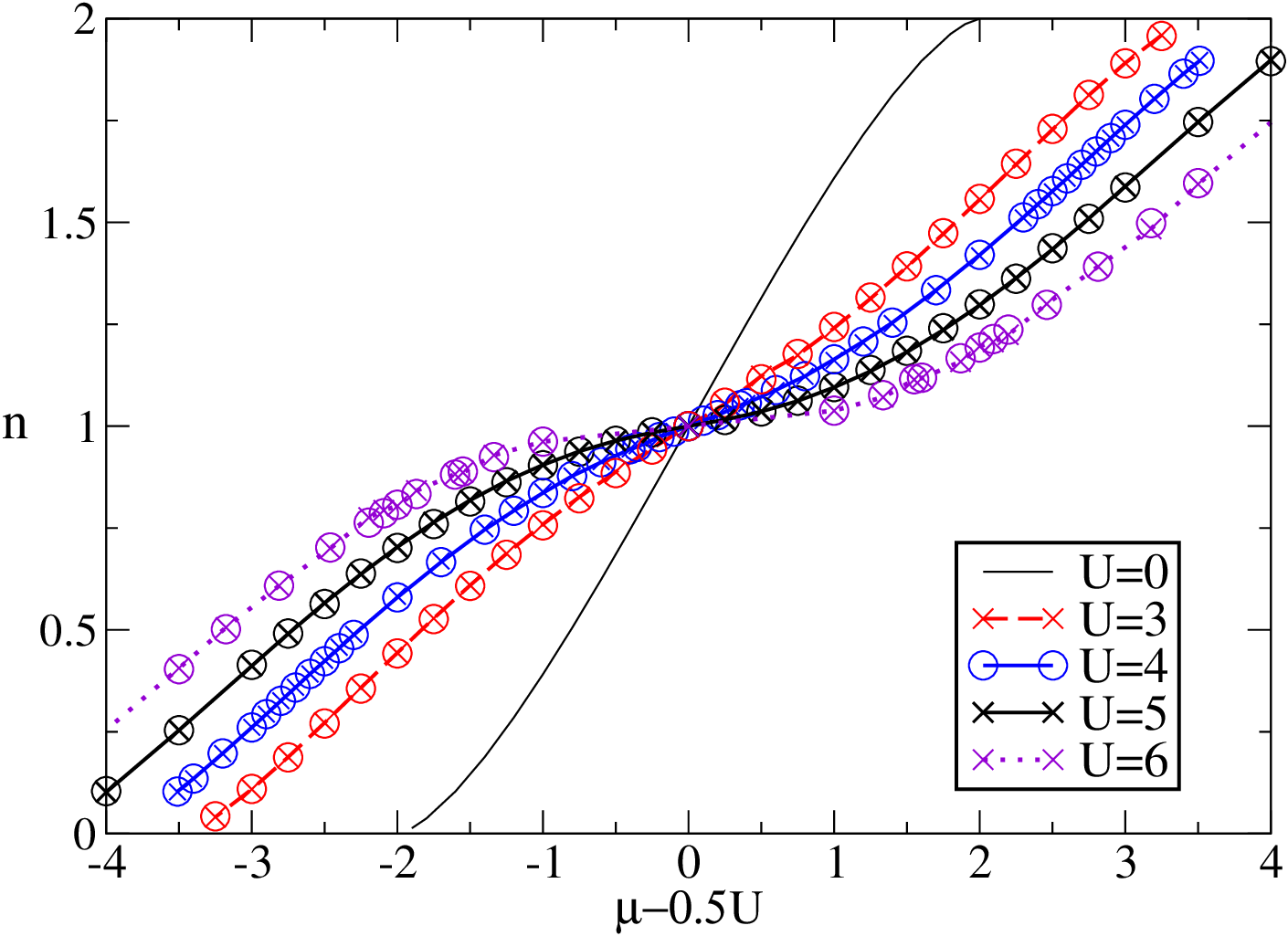}
\caption{The occupation number $n$ as a function of $\bar\mu=\mu-0.5U$
for $U=0,3,4,5,6$, as calculated directly from the DMFT (circles) and from the
NRG fixed point(crosses). The flattening of the curve in the region $n\sim 1$
for the larger value of $U$ indicates the strong correlation regime. }
\label{occ}
\vspace*{0.5cm}
\end{center}
\end{figure}

 We can check the relation
in Eqn. (\ref{noccex}) for the occupation number $n$ by comparing the
values deduced by substituting the results for  $\tilde \mu$ and
$\tilde D$ into (\ref{noccex}) with those deduced  from a direct
evaluation of the expectation value of $n$ in the ground state.
The results are plotted Fig. \ref{occ} as a function of 
$\bar\mu$ for $U=3,4,5,6$. The  occupation number $n$
for the non-interacting case $U=0$ is shown for comparison.
 The values calculated from  Eqn. (\ref{noccex}) (crosses) and by  direct NRG
calculation (circles) can be seen to be in excellent agreement (within about 1\%). 
If we assume the relation,  $\tilde
n_\sigma=n_\sigma$, then the agreement can alternatively be regarded as a check on
the calculation of the renormalized parameters,   $\tilde\mu$ and
$\tilde D$. The effects of strong correlation  leading to a plateau region at the point of half-filling
are also evident in this plot.
\par

\bigskip
\bigskip

\noindent

\begin{figure}[!htbp]
\begin{center}
\includegraphics[width=0.45\textwidth]{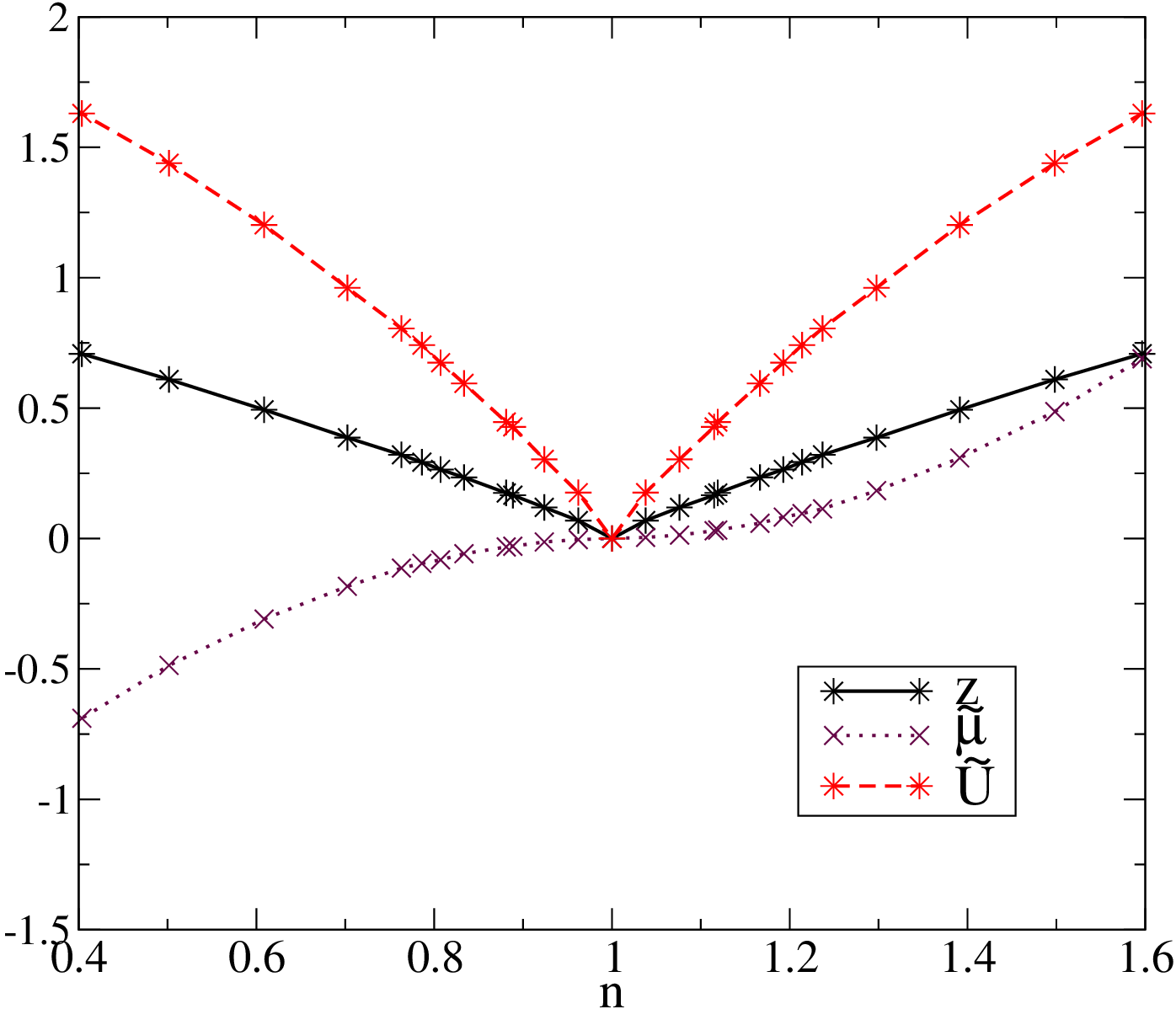}

\caption{ The quasiparticle weight
  $z=\tilde D/D$, the renormalized chemical potential $\tilde\mu$ and the on-site
  quasiparticle interaction $\tilde U$  for the
  model as a function of the occupation number $n$ for $U=6.0$. }
\label{rp2}
 \vspace*{0.5cm}
\end{center}
\end{figure}

\begin{figure}[!htbp]
\begin{center}
\includegraphics[width=0.45\textwidth]{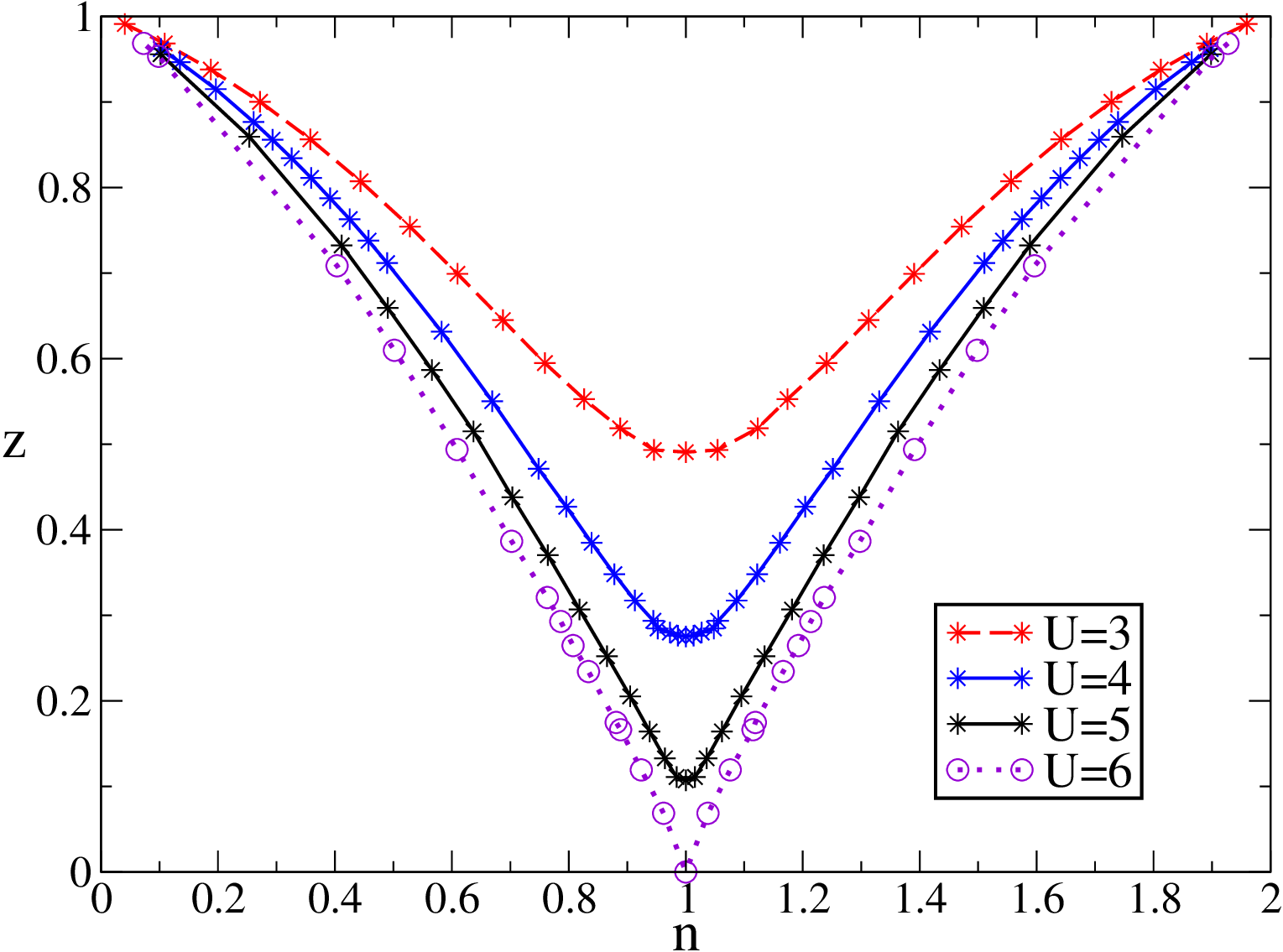}
\caption{The quasiparticle weight factor $z$ as a function of the occupation
  number $n$ for $U=3,4,5,6$.  }
\label{z_vs_n}
\vspace*{0.5cm}
\end{center}
\end{figure}
\noindent

\begin{figure}[!htbp]
\begin{center}
\includegraphics[width=0.45\textwidth]{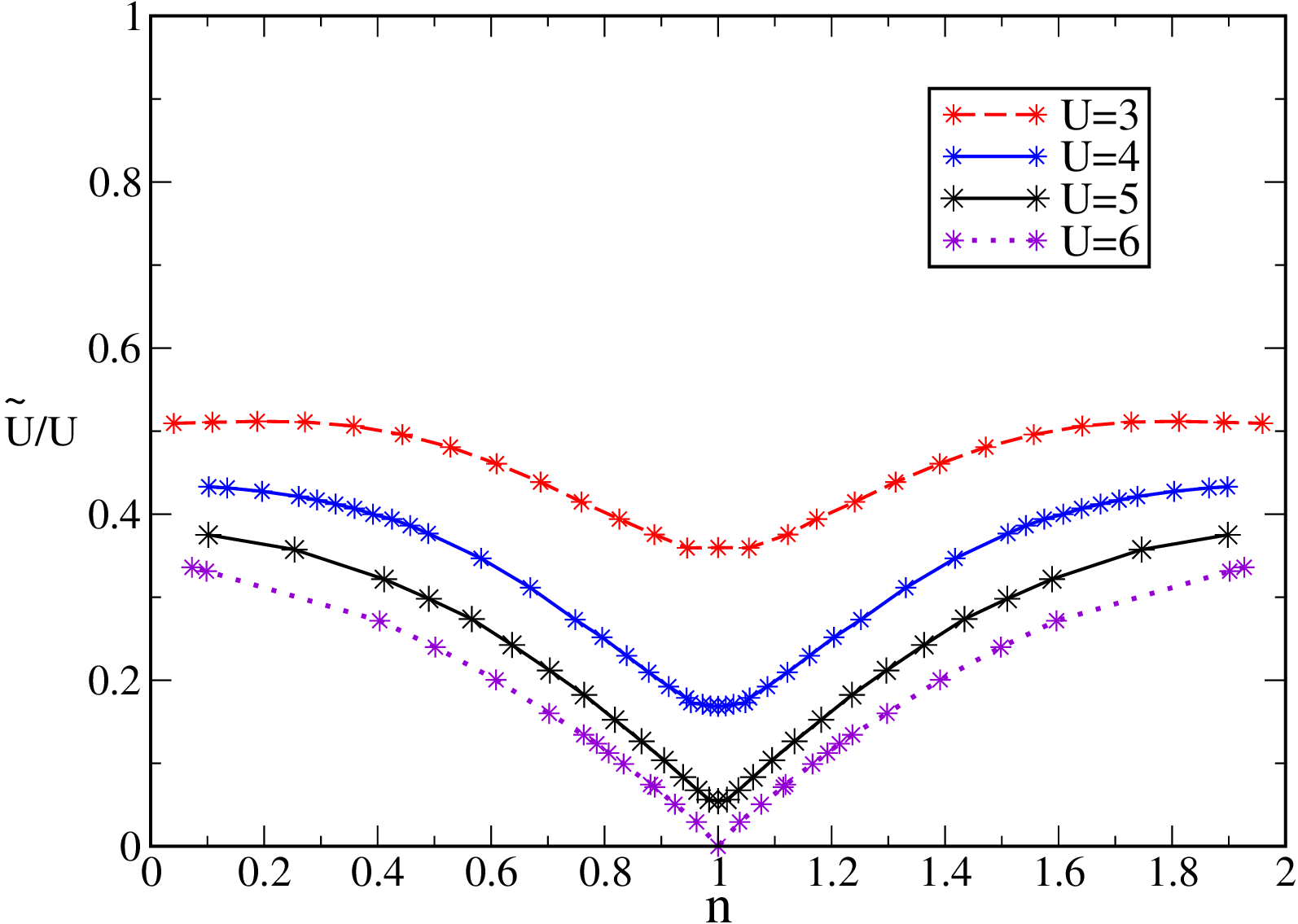}
\caption{The ratio of $\tilde U/U$ as a function of the occupation
  number $n$ for $U=3,4,5,6$. It can be seen that there is still some significant
renormalization of this quantity in the low particle density ($n\to 0$) and low hole
density regimes ($n\to 2$). }
\label{tildeU_vs_n}
\vspace*{0.5cm}
\end{center}
\end{figure}
\noindent
In Fig. \ref{rp2} we give show the results for $\tilde U$, $z=\tilde D/D$
and $\tilde \mu$ as a function of the filling factor $n=\sum_\sigma
n_\sigma$ for a value of $U=6$. As this value of $U$ is greater than 
$U_c$, the critical value for the Mott transition at half-filling, $z\to 0$
as the limit of half-filling $n\to 1$ is approached. We also find the $\tilde
U\tilde\rho_{\sigma}$ tends to the same value $\sim 0.815$ as $n\to 1$, so that
the values are independent of whether we approach the critical point for the
Mott transition by increasing $U$ at half-filling or with $U>U_c$ and letting
$n\to 1$. The renormalized quasiparticle chemical potential $\tilde\mu$ is
negative and approaches zero as $n\to 1$.
\par

 \begin{figure}[!htbp]
\begin{center}
\includegraphics[width=0.45\textwidth]{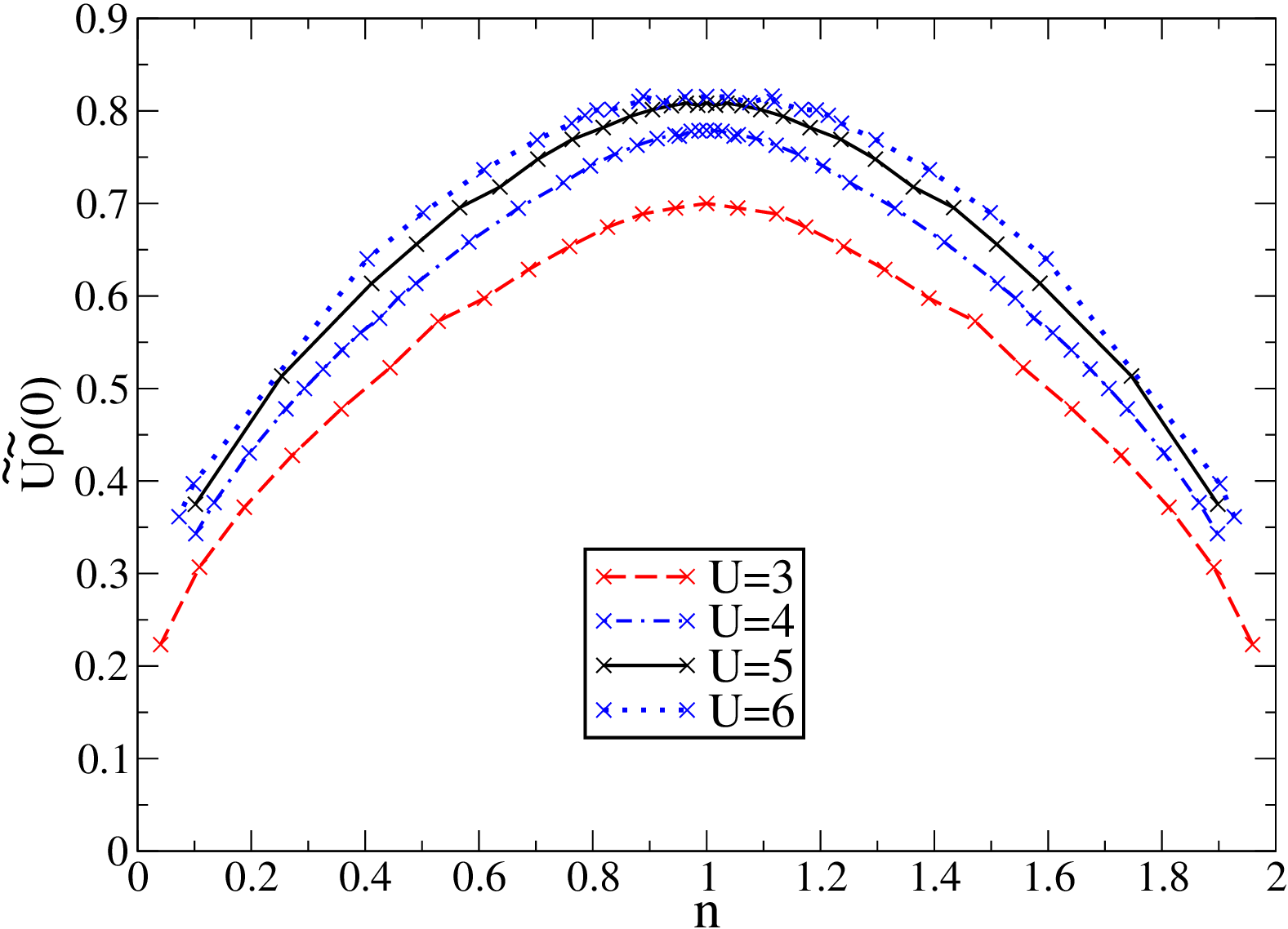}

\caption{ $\tilde U\tilde\rho(0)$ as a function of $n$ for $U=3,4,5,6$.  }
\label{tiUxtilrho_vs_n}
 \vspace*{0.5cm}
\end{center}
\end{figure}
\noindent

In Fig. \ref{z_vs_n} and Fig. \ref{tildeU_vs_n} we plot the  quasiparticle weight factor $z$
and the ratio $\tilde U/U$ as a function of the filling factor $n$. There is marked
minimum in both curves at the half-filling point, which is more pronounced for
the larger value of
$U$. If these are compared
with those for the Anderson impurity model\cite{HBK06} it can be
seen that there is a significant difference in the behaviour of $\tilde U/U$
in the regimes $n\to 0$ and $n\to 2$.  In the impurity case  $\tilde U/U\to 1$
so that the renormalization effects are negligible in these limits, whereas
 for the Hubbard model there is still some significant renormalization due to
 the phase space available for scattering. This can be estimated following
 Kanamori \cite{Kan63}, who calculated an effective interaction $U_{\rm eff}$,
using perturbation theory for the lattice model, taking  into account the renormalization due to
 repeated particle-particle scattering, which is the dominant process in the 
low density limit.      
This calculation takes the form,
\begin{equation} U_{\rm eff}={U\over 1-U\Pi^{p,\uparrow}_{p,\downarrow}(0)},
\end{equation}
where the particle-particle propagator $\Pi^{p,\uparrow}_{p,\downarrow}(0)$
at zero frequency in the low density limit  is given by  
\begin{equation}\Pi^{p,\uparrow}_{p,\downarrow}(0)=\int_{-D}^{D}\int_{-D}^{D
}{D(\epsilon,-D)D(\epsilon',-D)\over
(\epsilon+\epsilon')} d\epsilon d\epsilon'.\label{prop}\end{equation}
 The evaluation of (\ref{prop}) using the density of states given
in Eqn. (\ref{dos}) for $D=2$,   gives 
 $\Pi^{p,\uparrow}_{p,\downarrow}(0)=-0.3023$. The results for $ U_{\rm
   eff}$  are then  $U_{\rm eff}/U=0.524,\,0.453,\,0.398,\,0.355$,
for $U=3,4,5,6$.  We can identify  $ U_{\rm eff}$ as   $\tilde U$ in the
low density regime. From the results given in  Fig. \ref{tildeU_vs_n} we estimate these 
as    $\tilde
U/U=0.51,\,0.44,\,0.37,\,0.34$ for $U=3,4,5,6$ respectively. These are 
clearly in general agreement with  the Kanamori estimate, slightly
smaller but by less than 5\% difference in all cases. The quasiparticle weight factor $z$ in the lattice case does approach 
unity as  $n\to 0$ and $n\to 2$ as in the impurity case. \par

In Fig. \ref{tiUxtilrho_vs_n} we plot the dimensionless product  $\tilde
U\tilde\rho(0)$ which gives a measure of relative the strength of the
on-site quasiparticle interaction. For the single impurity Anderson model
in the Kondo limit  $\tilde
U\tilde\rho(0)\to 1$.  For the Hubbard model it can be seen to increase
steadily on the approach to the most strongly correlated situation at
half-filling. As noted earlier in the approach to the Mott transition,
either by increasing $U\to U_c$ at half-filling or as $n_d\to 1$ for $U>U_c$,
we get the same limiting value   $\tilde
U\tilde\rho(0)\to 0.815$. Almost the same limiting value has been obtain
for this quantity in studies of the Hubbard-Holstein model both on the
approach to the Mott transition and also in the localized limit due to
bipolaron formation\cite{BH10}. In the impurity case the result   $\tilde
U\tilde\rho(0)\to 1$  could be deduced from the condition that the charge
susceptibility tends to zero in the strong correlation regime. For the Hubbard model  we do not have an exact result for the charge susceptibility 
in terms of renormalized parameters to see if a similar argument could be used to
deduce
the limiting value of $\tilde
U\tilde\rho(0)$ on the approach to the Mott transition. \par  

\section{Static Response Functions}

If we express the zero temperature  static response function $\chi$  in the form,
\begin{equation}
\chi=\tilde\eta\tilde\chi^0,
\end{equation}
where $\tilde\chi^0$ is the corresponding function evaluated for the
renormalized but non-interacting quasiparticles, then the coefficient
$\tilde\eta$, is a dimensionless quantity and a measure of the effect of the quasiparticle interactions.
In the non-interacting case $U=0$,  $\tilde\eta=1$ as $\tilde\chi^0=\chi$.
On the approach to a quantum critical point, if the non-interacting 
quasiparticle susceptibility $\tilde\chi^0$  diverges, the corresponding
susceptibility $\chi$ will also diverge if $\tilde\eta$ tends to a finite
limit as $z\to 0$. However, not all susceptibilities will  be expected to
diverge at the transition point, so if  $\chi$ remains finite or zero as $z\to 0$ and
 $\tilde\chi^0$ diverges, then we require $\tilde\eta\to 0$.

\begin{figure}[!htbp]
\begin{center}
\includegraphics[width=0.45\textwidth]{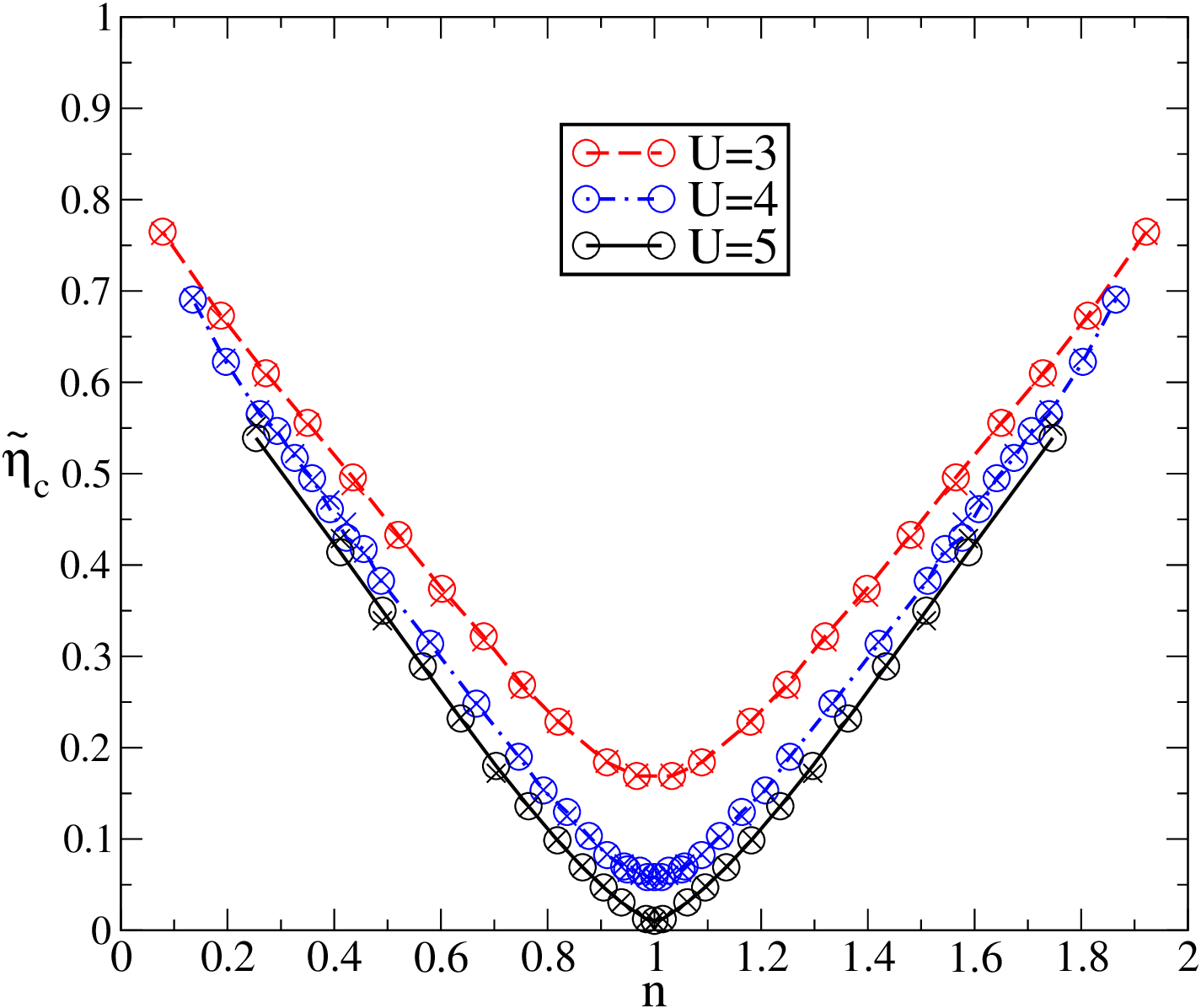}
\caption{$\tilde\eta_c=\chi_c/\tilde\chi_c^{(0)}$, where $\chi_c$ is the uniform charge susceptibility, plotted as a function of $n$ for $U=3,4,5$.
 }
\label{eta_c_latt}
\vspace*{0.5cm}
\end{center}
\end{figure}

We can deduce an expression for the zero temperature uniform charge
susceptibility $\chi_c$  by differentiating
Eqn. (\ref{noccex}).  The susceptibility for the non-interacting 
quasiparticles in this case given by $\tilde\chi^0=2\tilde\rho(0)$, and
$\tilde\eta_c$  by
 \begin{equation}
  \tilde\eta_c =z{d(\tilde\mu/z)\over d\mu}.
\label{chic}
 \end{equation}
The coefficient $\tilde\eta_c$ deduced from Eqn. (\ref{chic}) using the renormalized
 parameters is plotted 
 in Fig. \ref{eta_c_latt} (crosses) as a function of the site occupation value
$n$ for  $U=3,4,5,6$. The values of  $\tilde\eta_c$ can alternatively  be deduced from  $\chi_c$ by
taking  the derivative of the  occupation
number $n$, as calculated from the NRG ground state, with respect to $\mu$, and
 dividing the result by $2\tilde\rho(0)$.
 The results of this calculation are shown as circles in
 Fig. \ref{eta_c_latt}. 
We note that in the low density limit of electrons $n\to  0$, and the
 corresponding limit for holes $n\to  2$, that the values of  $\tilde\eta_c$
would appear to be lower than that for the `bare' electrons or holes
 $\tilde\eta_c =1$, even though $z\to 1$ in these limits. This must be due to
fact that there is  phase space available for the  particle-particle scattering
that led to a renormalization of $\tilde U$ from the bare value in these limits.

There is a steady decrease in  $\tilde\eta_c$ from the values
at  $n\sim 0$ and $n\sim 2$
to a minimum
at half-filling. The value $\tilde\eta_c$ at the half-filling is already very small for $U=5$
 and goes zero at the transition $U=5.98$. As $\tilde\rho(0)$ 
diverges on the approach to  the transition point this implies that the
charge susceptibility is either finite or zero in this limit. The fact that
the occupation number $n$  versus $\bar\mu$  as shown in Fig. \ref{occ} 
becomes flat for $U<U_c$, and there is a discontinuous jump in the values
of $\bar\mu$ between $n\to 1-$ and $n\to 1+$, means that $\chi_c\to 0$
as $U\to U_c$.

\begin{figure}[!htbp]
\begin{center}
\includegraphics[width=0.45\textwidth]{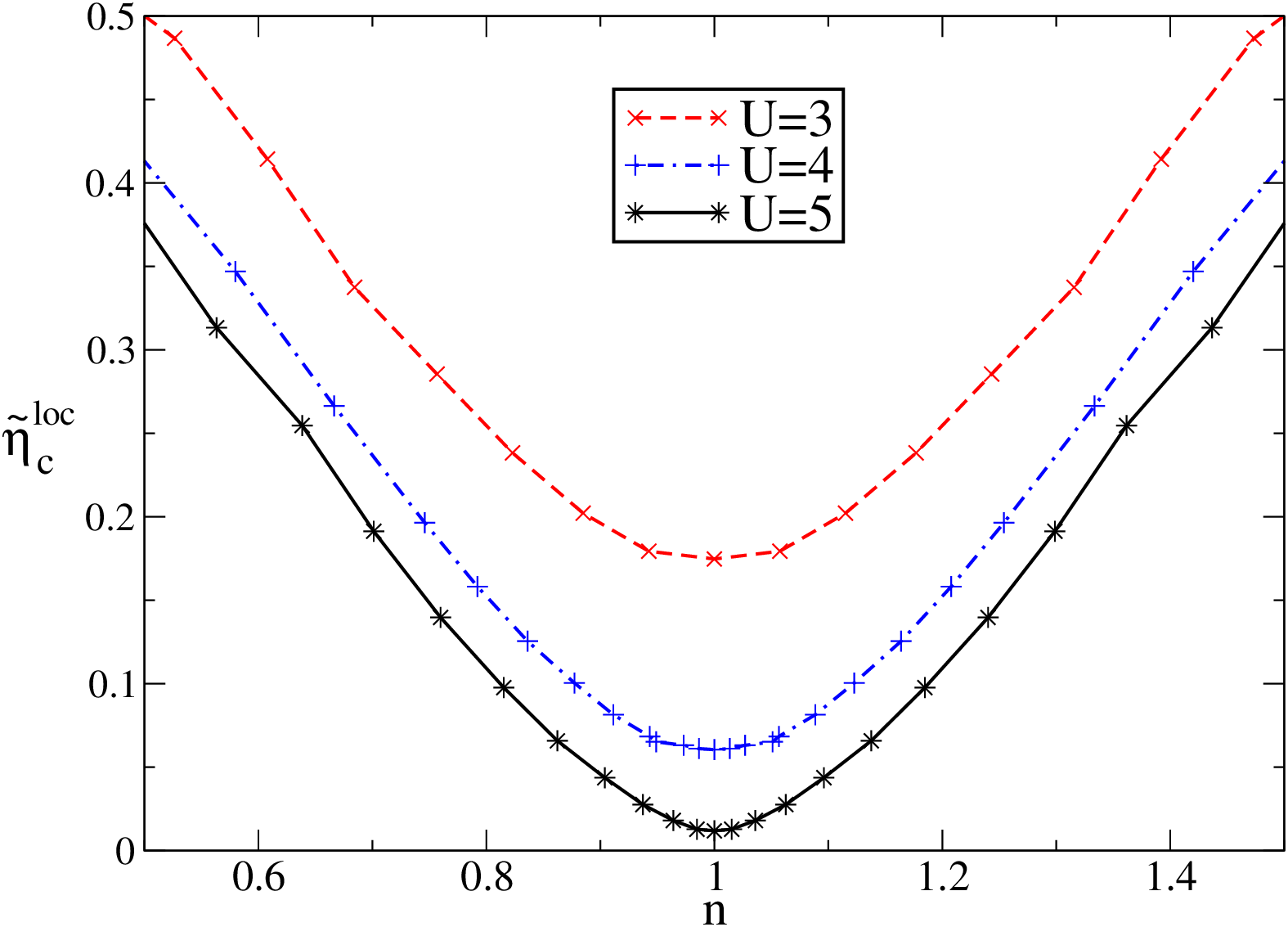}
\caption{$\tilde\eta^{\rm loc}_c=\chi^{\rm loc}_c/2\tilde\rho(0)$, where $\chi^{\rm loc}_c$ is the local charge susceptibility, plotted as a function of $n$ for $U=3,4,5$.
 }
\label{eta_c_local}
\vspace*{0.5cm}
\end{center}
\end{figure}

From the NRG results we can calculate the local on-site dynamic charge susceptibility $\chi_c^{\rm loc}(\omega)$ at $\omega=0$, which we will denote
by  $\chi_c^{\rm loc}$. We can define a coefficient $\eta_c^{\rm loc}$
via the relation,   $\chi^{\rm loc}_c =2\tilde\eta^{\rm loc}_c\tilde\rho(0)$.
 The values of $\tilde\eta^{\rm loc}_c$ deduced from the NRG results  are  shown as a function of the occupation number $n$ in 
Fig. \ref{eta_c_local}. The results and general trend are very similar to
 those for that uniform charge susceptibility shown in Fig. \ref{eta_c_latt}.
\par

\begin{figure}[!htbp]
\begin{center}
\includegraphics[width=0.45\textwidth]{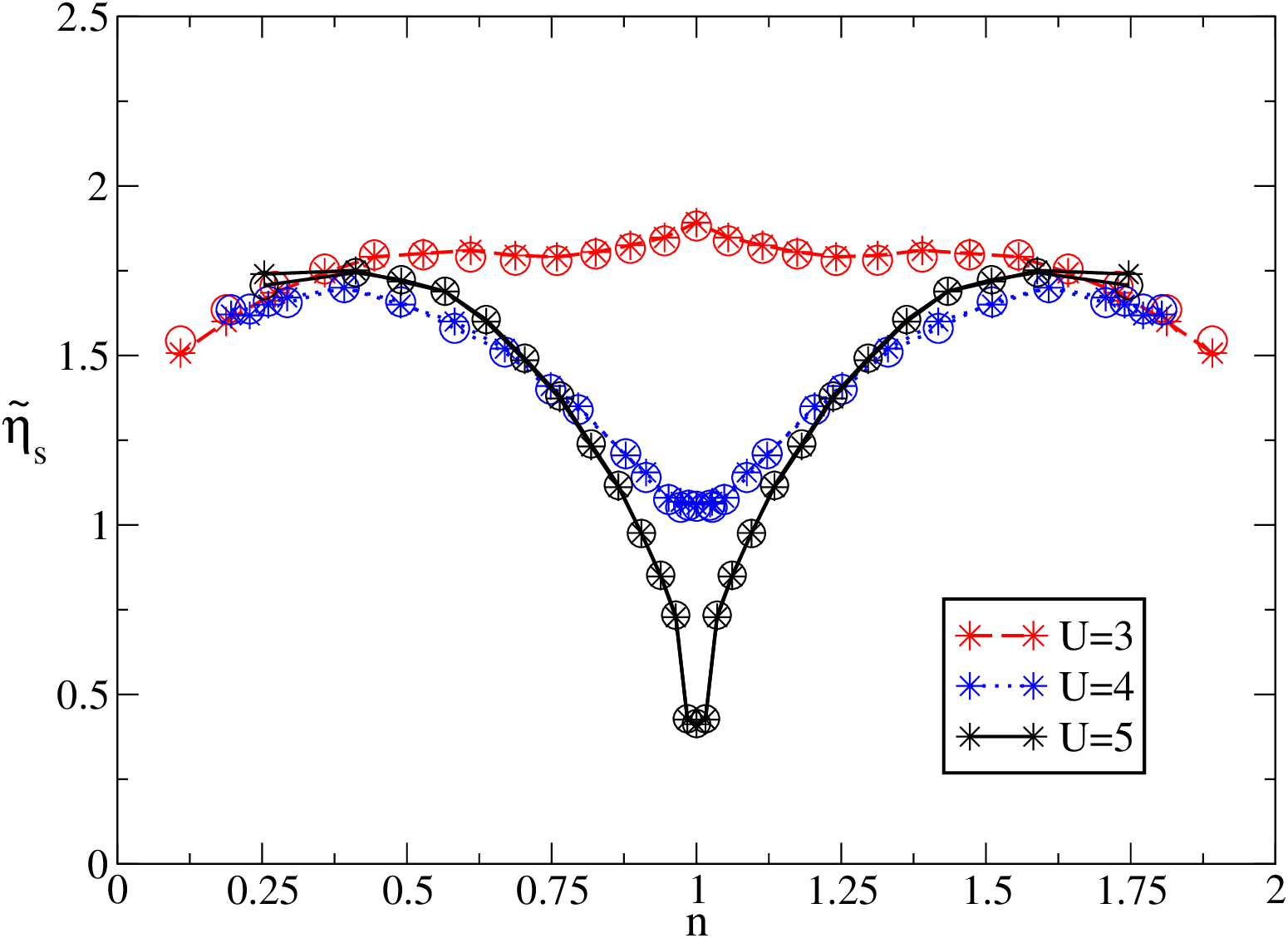}
\caption{$\tilde\eta_s=\chi_s/\tilde\chi_s$, where $\chi_s$ is the uniform spin susceptibility, plotted as a function of $n$ for $U=3,4,5$.
 }
\label{eta_s}
\vspace*{0.5cm}
\end{center}
\end{figure}

We find distinct differences, however, between the local and uniform susceptibilities in
the case of the spin. The zero field uniform susceptibility at $T=0$ can be expressed in the form,
 \begin{equation}
   \chi_s ={1\over 2}(g\mu_{\rm B})^2\tilde\eta_s\tilde\rho(0),\quad{\rm where}\quad
\tilde\eta_s
={\rm lim}_{h\to 0}{{(\tilde \mu_{\uparrow}-\tilde \mu_{\downarrow})}\over 2h},
 \label{chis}
 \end{equation}
where the factor $\tilde\eta_s$ is due to the interaction between the
quasiparticles, and is equivalent to the usual definition of the Wilson
$\chi/\gamma$ ratio. It  can be calculated from Eqn. (\ref{chis}) using the
results for the renormalized parameters in a magnetic field. Alternatively it
can be deduced from the magnetization  $m(h)$ calculated from the NRG ground state  using  $\tilde \eta_s=\lim_{h\to 0}m(h)/h\tilde\rho(0)$.
The results for  $\tilde\eta_s$ are shown in Fig. \ref{eta_s} as a function of $n$ for 
$U=3, 4, 5$ and  $6$. The points marked with a cross indicate those calculated
from the renormalized parameters, and those with circles are deduced from the
NRG magnetization. The two sets of results are in good agreement.
 There is a marked change in the form of $\tilde\eta_s$ on the approach to
 half-filling as the value of $U$ is increased from 3 to 5. For $U<4$
$\tilde\eta_s>1$, there is  an enhancement of the quasiparticle
susceptibility due to the quasiparticle interactions, increasing from the low
density regime with a slight peak at half-filling. There is also an
enhancement for $U=4$ in the low density regime but a significant dip on the
approach to half-filling where it has a minimum with $\tilde\eta_s\sim 1$. The same
trend can be seen for the case $U=5$ but the dip at half-filling is much much
greater and such that $\tilde\eta_s <1$. This means that the quasiparticle
interactions are tending to suppress rather than enhance the free quasiparticle
susceptibility, which was also found in the calculation of Bauer\cite{Bau09a}.
 Such a suppression would be expected from an antiferromagnetic
interaction between the quasiparticles. For large $U$ in the localized limit
at half-filling the Hubbard model can be mapped into an antiferromagnetic
Heisenberg model and  has an antiferromagnetic ground state, so the
quasiparticle interactions could be precursors of this limit. It would be
interesting to calculate  $\tilde\eta_s$  near half-filling for values of
$U$ on the approach to the Mott transition $U\to U_c$. Unfortunately 
it  becomes very difficult in this regime to achieve self-consistency of the
the DMFT equations in very weak magnetic fields in this regime,
such that numerically we can make no reliable predictions for the behavior
of $\eta_s$ as $U\to U_c$. However, there is an interesting analogy
with a two quantum dot model with an antiferromagnetic interaction
between the dots, which has a quantum critical point. In that case,
though the quasiparticle weight $z\to 0$ on the approach to the critical point
the uniform susceptibility remains finite\cite{NCH12a,NCH12b}. This implies 
 $\tilde\eta\to 0$ as $z\to 0$. We speculate the something similar might
hold in this case also, and the trend seen in Fig. \ref{eta_s} with increasing
$U$ will be such that the value of  $\tilde\eta_s$ will dip to zero at half-filling
as $U\to U_c$. Further evidence to test whether this might be the case could be derived from
a calculation of the zero field susceptibility to higher order $\tilde U$ in the
RPT, along the lines used in Ref.~\onlinecite{PH15b}, and this is under active consideration. The results could  also be tested against those deduced from the NRG for a range of values of $U$. \par

\begin{figure}[!htbp]
\begin{center}
\includegraphics[width=0.45\textwidth]{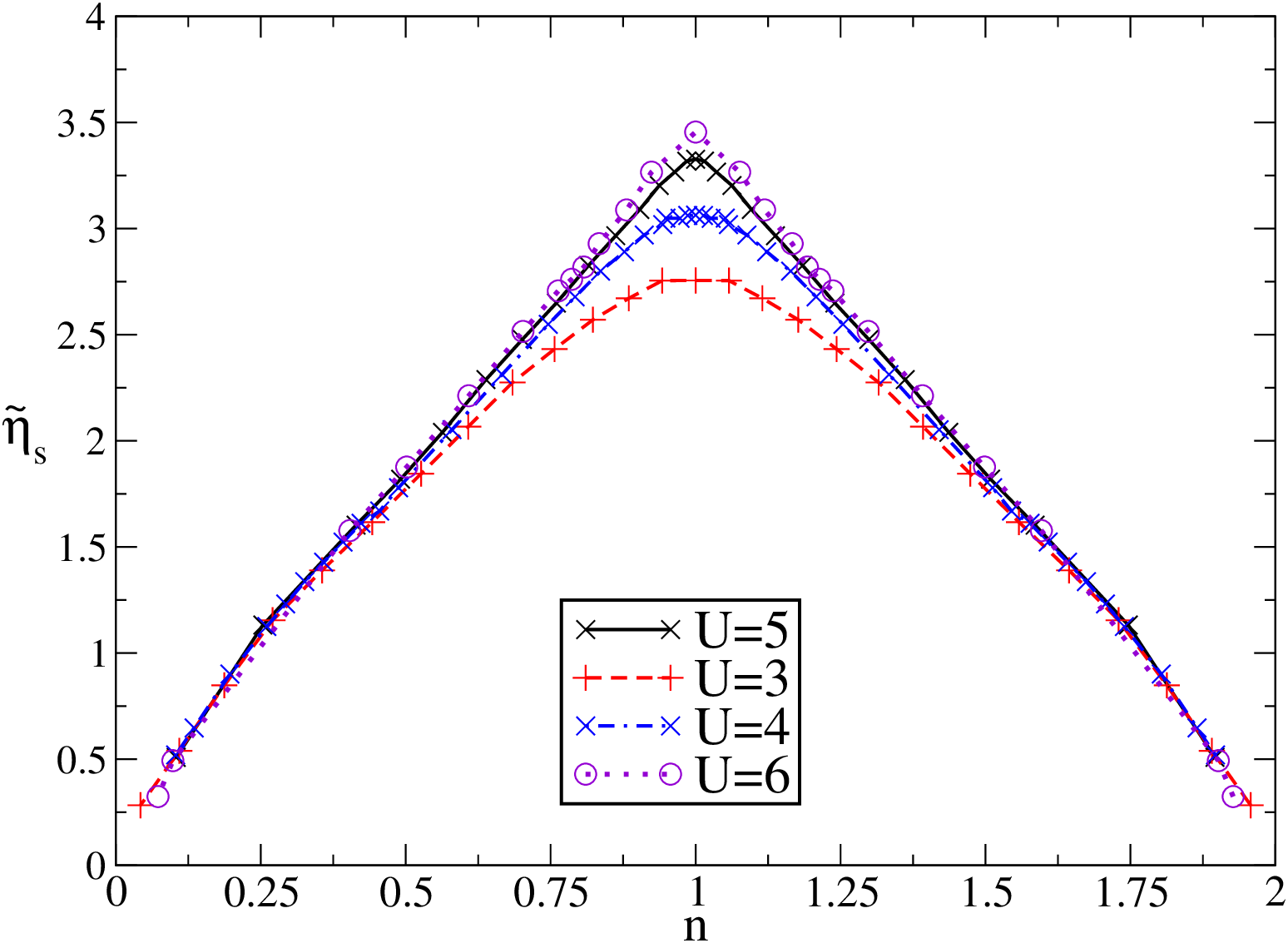}
\caption{ $\tilde\eta^{\rm loc}_s=2\chi^{\rm loc}_s/\tilde\rho(0)$, where $\chi^{\rm loc}_s$ is the local spin susceptibility, plotted as a function of $n$ for $U=3,4,5,6$.
 }
\label{etas_local}
\vspace*{0.5cm}
\end{center}
\end{figure}

The local spin susceptibility has a completely different
behavior on the approach to half-filling. 
We define  $\chi_s^{\rm loc}$
as the $\omega=0$ value of the on-site spin correlation function  $\chi_s^{\rm
  loc}(\omega)$ which can be calculated using the NRG. We can define an
$\eta_s^{\rm loc}$  via the relation,  $\chi_s^{\rm loc} ={1\over 2}(g\mu_{\rm B})^2\tilde\eta_s^{\rm loc}\tilde\rho(0)$,
Results for
$\eta_s^{\rm loc}$ are shown for $U=3,4,5,6$ in Fig. \ref{etas_local}.
They all show a steady increase  on the approach to half-filling  to a finite maximum value at $n=1$.
There is only a significant difference between the results for the different values of $U$ in the region
near half-filling, the values for larger $U$ being larger.
\noindent

\noindent

\noindent

\noindent

\section{Renormalized Self-Energy Calculations \label{s4}}
Having deduced from the NRG the renormalized parameters $\tilde\mu$
and $\tilde D$,
which define the free quasiparticle density of states
$\tilde\rho_{\sigma}(\omega)$, and the renormalized on-site
quasiparticle interaction $\tilde U$, 
from the NRG, we are now in a position to use them 
in the RPT to calculate the
renormalized self-energy $\tilde\Sigma_\sigma(\omega)$.
The perturbation theory can proceed exactly along the same
lines as the RPT for the standard single impurity Anderson model.
The free quasiparticle Green's function is $\tilde G^{\rm imp}_{0,\sigma}(\omega)$
is the propagator in the expansion which is formally in powers 
of $\tilde U$. The main difference from the usual perturbation theory
in powers of the bare parameter $U$ is that the parameter $\tilde U$
is already renormalized. As a consequence
 counter terms have to
be included to ensure that there is no overcounting of renormalization
effects. These are determined
from the conditions that $\tilde\Sigma_\sigma(0)=0$,
 $\tilde\Sigma'_\sigma(0)=0$, and that $\tilde U=\tilde
 \Gamma^{(4)}_{\uparrow,\downarrow}(0,0,0,0)=z^2
 \Gamma^{(4)}_{\uparrow,\downarrow}(0,0,0,0)$, where
$ \Gamma^{(4)}_{\uparrow,\downarrow}(\omega_1,\omega_2,\omega_3,\omega_4)$ is the
full local four-vertex. \par
To test the RPT results for $\tilde\Sigma_\sigma(\omega)$ in the low energy
regime 
against the NRG calculations for the self-energy $\Sigma_\sigma(\omega)$,
it will be convenient to use the relation between their imaginary
parts,
\begin{equation} {\rm Im}\,\Sigma(\omega)={1\over z} {\rm Im}\,\tilde\Sigma(\omega),
\end{equation}
which follows directly from the definition of the renormalized self-energy
$\tilde\Sigma_\sigma(\omega)$.\par
The lowest order correction term for ${\rm Im}\tilde\Sigma(\omega)$ is second
order in $\tilde U$. It has been shown for the  particle-hole
symmetric Anderson model  that this term gives the asymptotically exact result to leading order as $\omega\to 0$ and $T\to 0$,
 for all
values of $U$.  This result then enables one to calculate exactly the leading order temperature
dependence of the conductivity as $T\to 0$. Here we perform the same
calculation
using the parameters derived for the lattice and test the results with those
derived directly from the NRG. Working to second order in $\tilde U$ we can
use the standard perturbation theory to evaluate  $ {\rm
  Im}\tilde\Sigma(\omega)$. The two counter terms that ensure
$\tilde\Sigma_\sigma(0)=0$ and $\tilde\Sigma'_\sigma(0)=0$, to this order are
real and do not contribute
to the imaginary part of $\tilde\Sigma(\omega)$. There is also no counter term
correction to the condition $\tilde
U=\tilde\Gamma_{\uparrow,\downarrow}(0,0,0,0)$ to second order. 
We then find
\begin{equation} {\rm Im}\tilde\Sigma(\omega)=\pi\tilde U^2 
    \int\tilde\rho(\epsilon)\tilde\rho(\epsilon')\tilde\rho_(\omega-\epsilon-\epsilon')
F(\omega,\epsilon,\epsilon')d\epsilon d\epsilon',
\end{equation}
where
\begin{equation}F(\omega,\epsilon,\epsilon') =(1-f(\epsilon)-f(\epsilon'))
f(\epsilon+\epsilon'-\omega)+f(\epsilon)f(\epsilon').
\end{equation}
with $f(\epsilon)=1/(1+e^{\epsilon/T})$.
This leads to the asymptotic form  for small $\omega$ and $T$,

\begin{equation} {\rm Im}\,\Sigma(\omega)\sim
-{\pi\over 2z}\tilde\rho(0)^3\tilde U^2(\omega^2+(\pi T)^2).
\label{asymp}\end{equation}
If we introduce a renormalized energy scale via  $1/\tilde\rho(0)=4T^*$ 
(in an impurity model in the Kondo regime $T^*$ corresponds to the Kondo
temperature $T_{\rm K}$), then we can rewrite this expression in the form, 
\begin{equation} {\rm Im}\Sigma(\omega)=
-{\pi C^2\over 32 \rho(0)}\left[\left({\omega\over
      T^*}\right)^2+\left({\pi T\over T^*}\right)^2\right]+....
,\end{equation}
where $C=\tilde \rho(0)\tilde U$ is a dimensionless parameter. As mentioned
      earlier $C$ tends to the value $0.816$ in the approach to the Mott
      transition ($z\to 0$). As a consequence all the renormalized parameters 
can be expressed in terms of the single energy scale 
$T^*$ on the approach to the Mott transition. This same behavior was already found in a local model, which has
two types of zero temperature transitions, on the approach to each critical point\cite{NCH12a,NCH12b}.
As $4T^*=1/\tilde\rho(0)$, and at particle-hole symmetry $\tilde\rho(0)=2/\pi\tilde D$, then $T^*=z\pi D/8$, proportional to $z$ so this is also equivalent to the $\omega/z$ scaling found in Ref.~\onlinecite{ZHPMGS13}.
\par
\begin{figure}[!htbp]
\begin{center}
\includegraphics[width=0.45\textwidth]{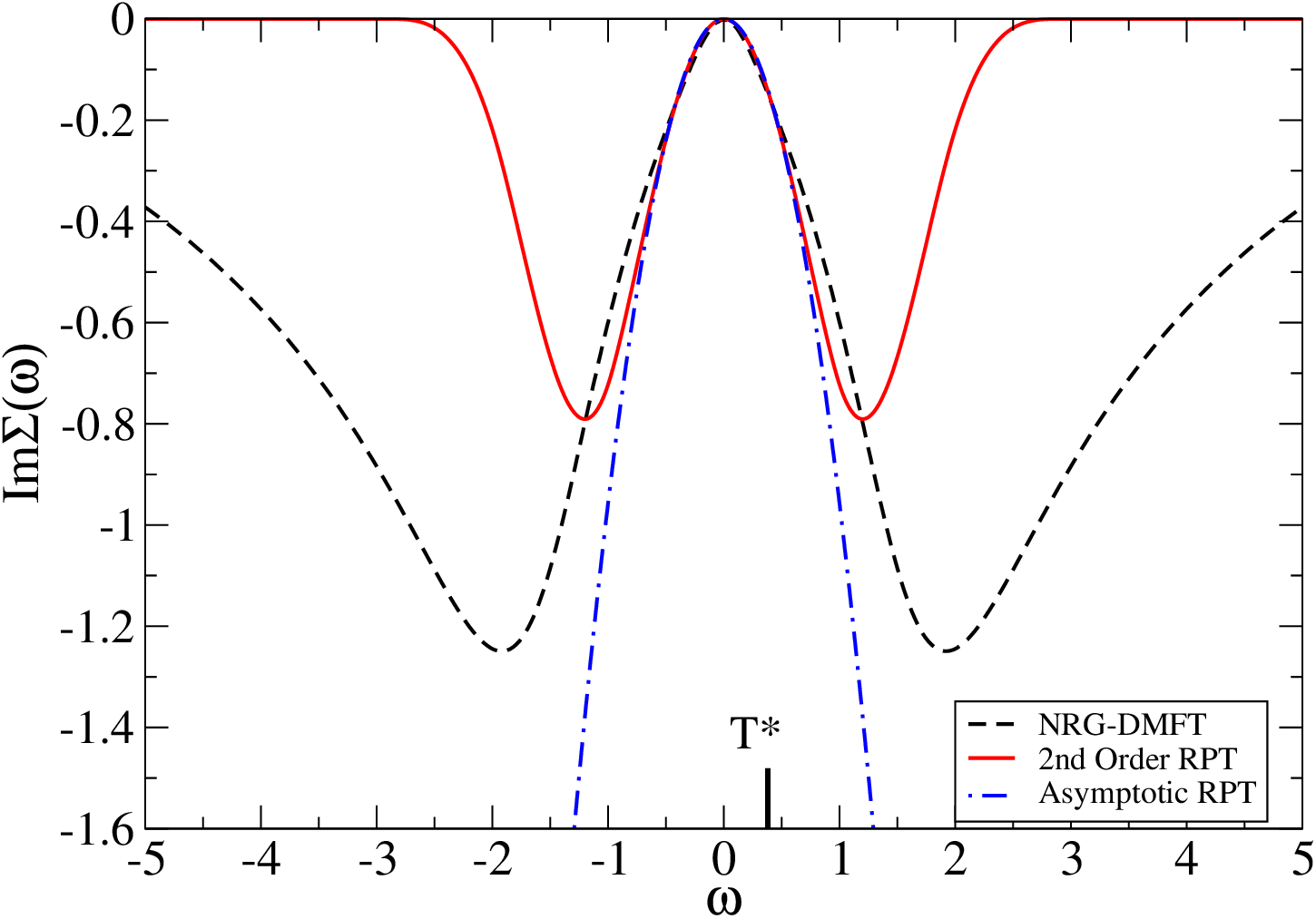}
\caption{
The imaginary part of the self-energy
for U=3.0, $T^*=0.38$,   compared with the corresponding NRG results.
 }
\label{isigu3.0}
\end{center}
\end{figure}

\begin{figure}[!htbp]
\begin{center}
\includegraphics[width=0.45\textwidth]{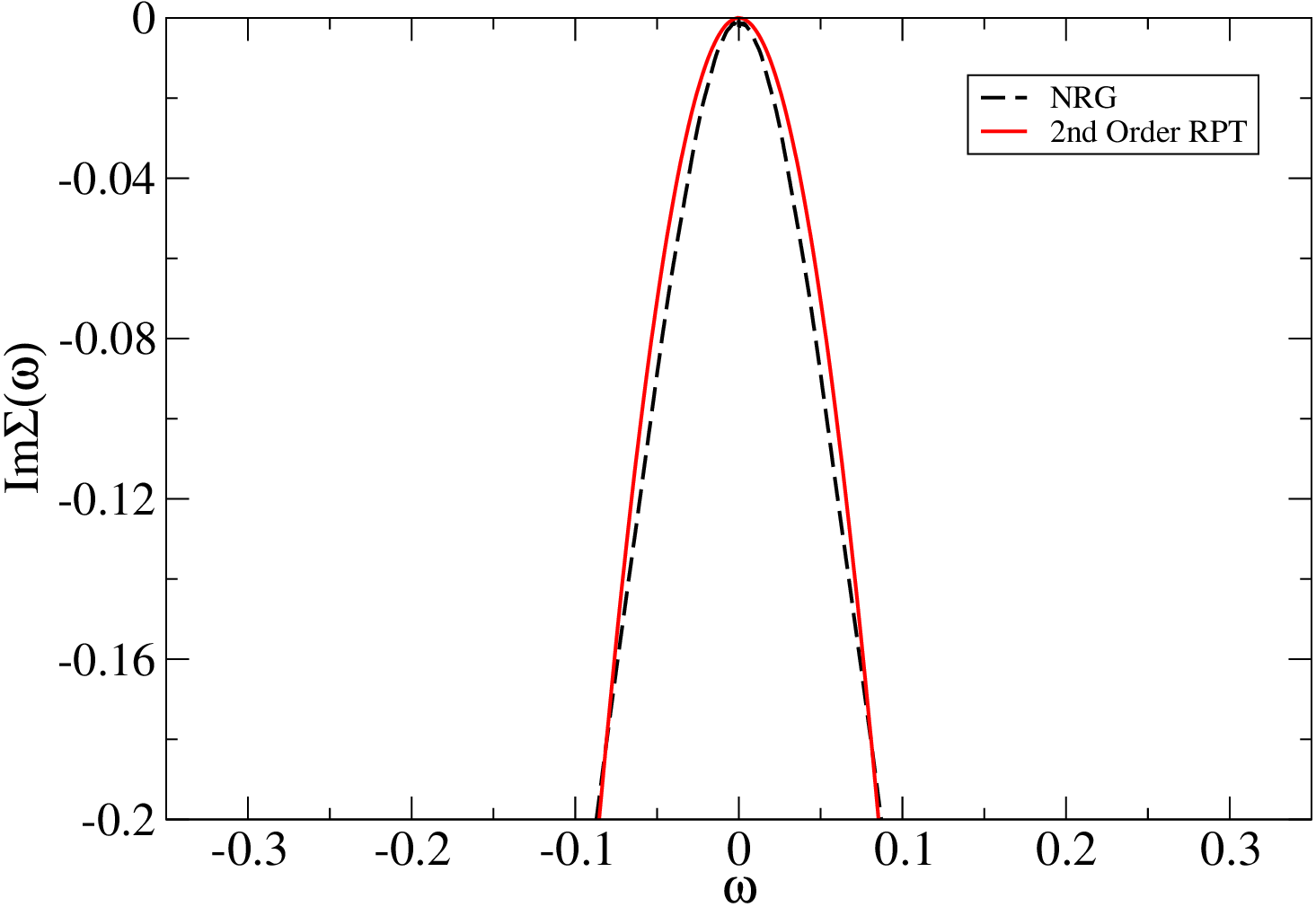}
\caption{The imaginary part of the self-energy
for U=5.0, $T^*=0.084$,  compared with corresponding NRG results.}
\bigskip
\label{isigu5.0}
\end{center}
 \end{figure}

  
We can check the predictions of the RPT for ${\rm Im}\Sigma(\omega)$ 
by making a  comparision with the results for this quantity
obtained directly from the NRG calculations.
In Fig. \ref{isigu3.0} we compare with the RPT and NRG results
for ${\rm Im}\Sigma(\omega)$ at
 $T=0$ for the half-filled model with $U=3$. The second order result
 clearly describes the behaviour over the low energy scale $|\omega|<T^*$.
Over this region there is very little difference between the full second order
result and the asymptotic result (\ref{asymp}). In Fig.  \ref{isigu5.0}
a similar comparison is made between the NRG and asymptotic result
for a larger value $U=5.0$. Again there is good  agreement  over the range
 $|\omega|<T^*$. It is difficult to make a comparison for larger values of
$U$ near the Mott transition as $T^*$ becomes very small as $T^*\to 0$
for $U\to U_c$. Due to the discrete spectrum used for the bath in the NRG
calculations, the spectra generated consist of sets of delta functions
which have to be broadened to give a continuous spectrum. This broadening
factor then introduces errors in determining the  coefficient of the
      $\omega^2$ term, which make it difficult to estimate reliably when
$T^*$ becomes very small.
\par
\begin{figure}[!htbp]
\begin{center}
\includegraphics[width=0.45\textwidth]{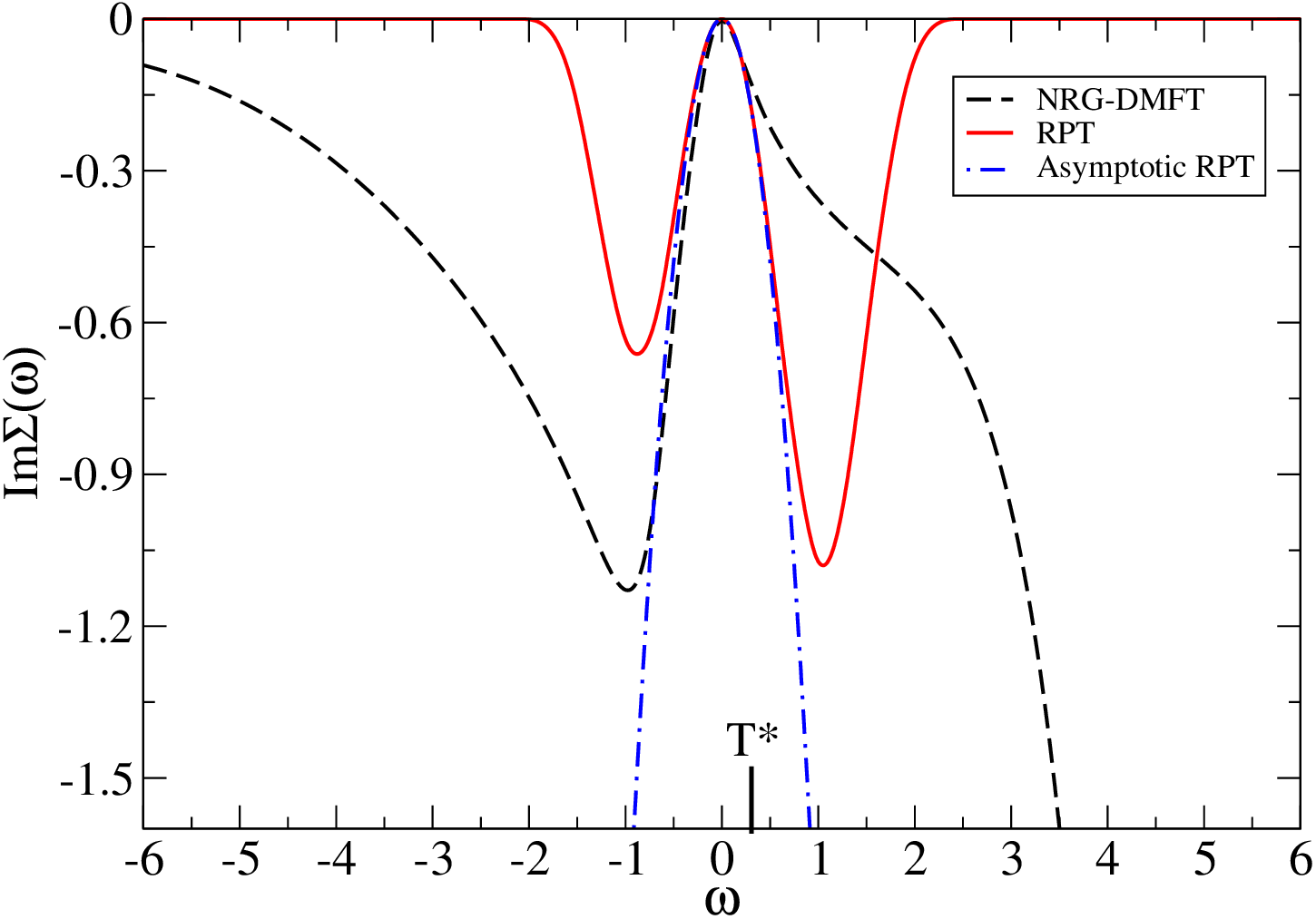}
\vspace*{0.5cm}
\caption{A comparison of the RPT result for the imaginary part of the self-energy for n=0.7, U=6,
$T^*\sim 0.31$ with the corresponding NRG-DMFT results. There is
 good agreement  for positive $\omega$ up to $\omega\sim T^*$
but the agreement extends to larger values of $|\omega|$ on the negative side.}
\label{isig_0.7_6.0}
\end{center}
\end{figure}

  In Fig. \ref{isig_0.7_6.0} we make a comparison of the results
in a case away from half-filling with $U=6.0$ and $x=0.7$. The agreement is again good
over the scale  $|\omega|<T^*$, but the NRG results deviates quite
markedly from the RPT second order result for $T^*<\omega < 2T^*$,
though it is still a good approximation for $-2T^*<\omega < -T^*$.\par
The indication from these results is that the second order RPT result
does lead to the correct asymptotic behaviour  for the imaginary part of the
 self-energy,  and so these results can be used to calculate the $T^2$
coefficient of the conductivity for this model. 
\vskip.3in

\noindent

\bigskip\bigskip
\vskip.2in

\bigskip

\section{Local Dynamic Response Functions \label{s5}}

\begin{figure}
\begin{center}
 \vspace*{0.5cm}
\includegraphics[width=0.40\textwidth]{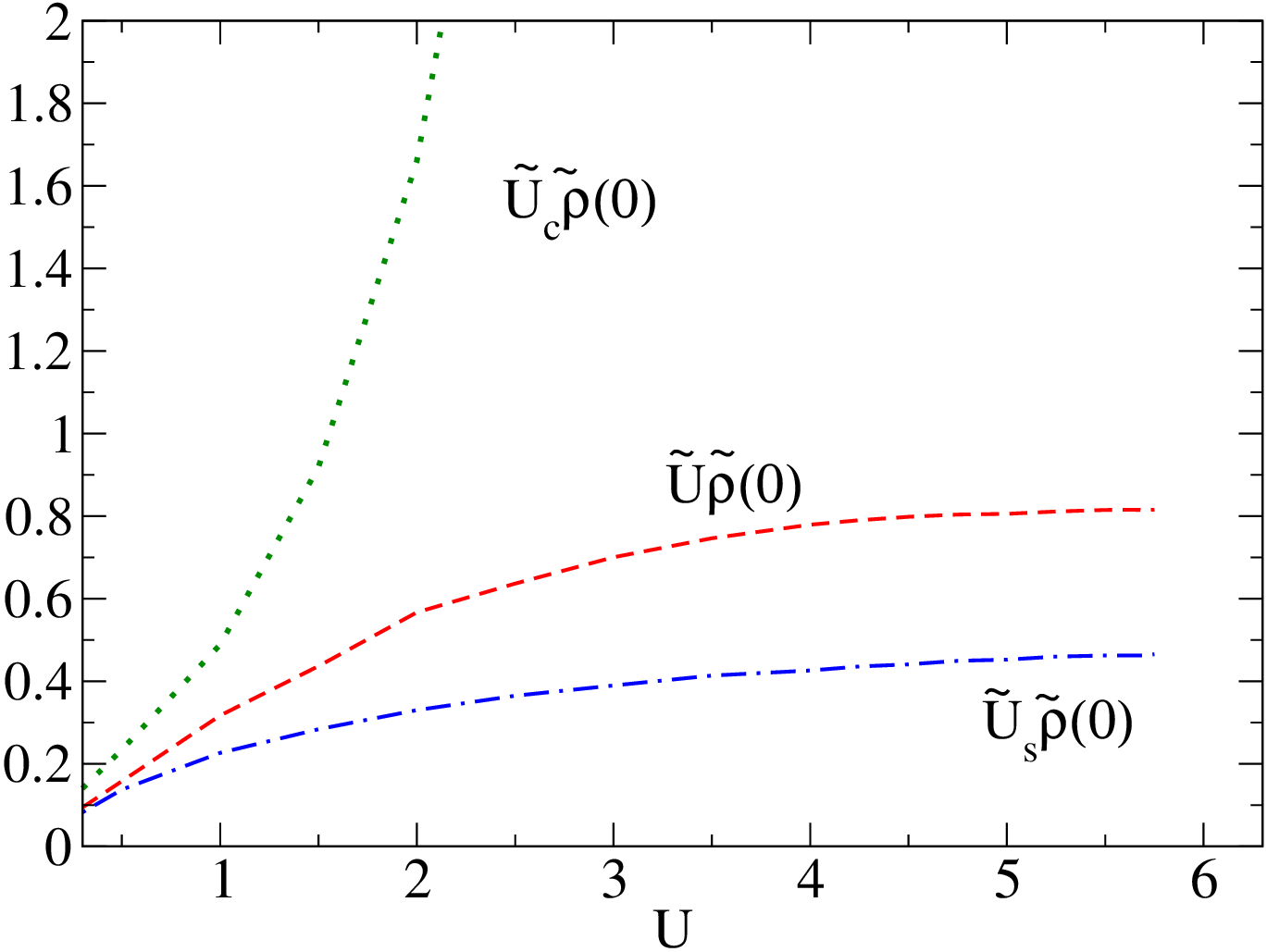}

\caption{ A plot  of $\tilde U\tilde\rho(0)$, 
 $\tilde U_s\tilde\rho(0)$ and 
 $\tilde U_c\tilde\rho(0)$ as a function of $U$ at half-filling. }
\bigskip
\label{rp_rho}
\end{center}
\end{figure}

\begin{figure}[!htbp]
\begin{center}
\includegraphics[width=0.45\textwidth]{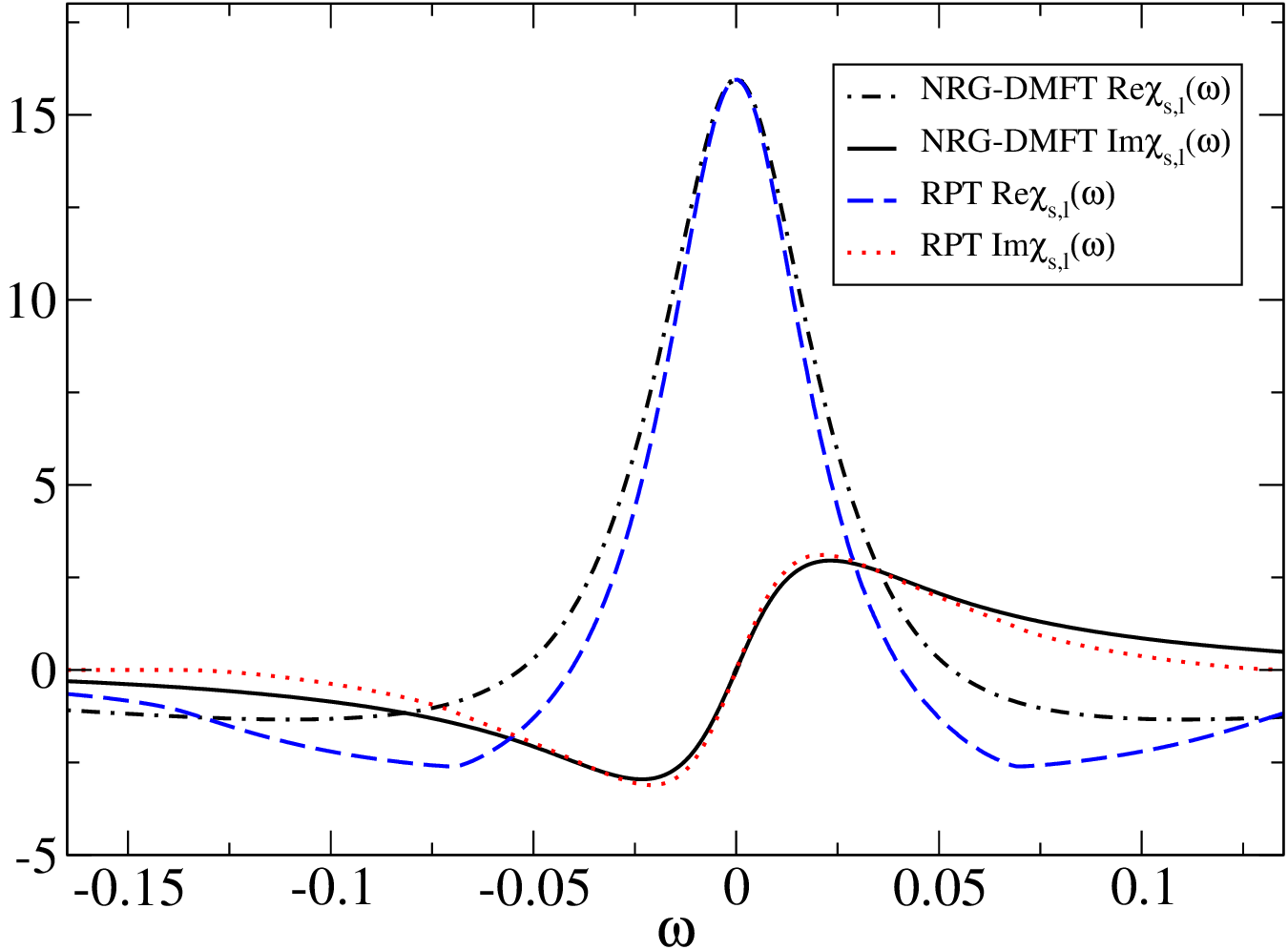}

\caption{
A comparison of NRG-DMFT results for $\chi_{s,l}(\omega)$ with the RPT
formula for U=5.6. }
\label{compchis}
\end{center}
\end{figure}

The calculations here proceed along similar lines for the effective
 impurity.  The equation for the transverse spin susceptibility is 
 \begin{equation}
 \chi_{s,t}(\omega)={\tilde\Pi^0(\omega)
 \over 1-\tilde
 U_{s}^{\rm loc}\tilde\Pi^0(\omega)},
 \label{rchit}
 \end{equation}
$\tilde U_s^{\rm loc}$ is the irreducible quasiparticle interaction in this
 channel and  $\tilde\Pi^0(\omega)$ is given by
 \begin{equation}
\tilde\Pi^0(\omega)= \int\int {f(\epsilon)-f(\epsilon')\over
  (\omega-\epsilon+\epsilon')}\tilde\rho(\epsilon)\tilde\rho(\epsilon')\, d\epsilon\, d\epsilon',
 \end{equation}
 where $\tilde\rho(\omega)$  is the free quasiparticles density of states given in Eqn.
 (\ref{qpdos}).
 In the absence of a magnetic field $\chi_{s, l}(\omega)$
is the same as the transverse response function apart from a factor 2, $\chi_{s,l}(\omega)=0.5 \chi_{s,t}(\omega)$.
The interaction term $\tilde U_{s}^{\rm loc}$ in the scattering channel is not the same as the on-site
 quasiparticle interaction $\tilde U$, calculated earlier, as the $\tilde U$,
 already includes some of these scattering terms for $\omega=0$, so $\tilde
 U_{s}^{\rm loc}=\tilde U -\lambda_3$, where $\lambda_3$ is the  counter term associated
 with the interaction. In the impurity case, assuming a flat wide band for the
 conduction electrons, it was
 possible
to derive an exact expression for $\chi_{s,l}(0)$, in terms of $\tilde U$,
 which enabled one to derive an explicit expression for $\tilde U^{\rm loc}_s$ in terms
 of
$\tilde U$. However,  the approximation of a flat wide band for the conduction
electron bath is not applicable to the effective impurity considered here,
so we need another way to estimate $\tilde U^{\rm loc}_s$. One possibility explored here is to
treat $\tilde U^{\rm loc}_s$ as a free parameter and use it to fit the value of (\ref{rchit}) at $\omega=0$,
as derived from the NRG-DMFT. We can then test how well the expression in 
Eqn. (\ref{rchit}) fits the NRG-DMFT results for the real and imaginary parts
of $\chi_{s,l}(\omega)$ as a function of $\omega$. In a similar way the local dynamic charge susceptibility
is $\chi_c(\omega)$ can be calculated from an expression of the same form as (\ref{rchit}) with $\tilde U^{\rm loc}_s$ replaced by $-\tilde U^{\rm loc}_c$.\par

The values of $\tilde U^{\rm loc}_s\tilde\rho(0)$ and $\tilde U^{\rm loc}_c\tilde\rho(0)$
deduced in this way for the model at half-filling are shown as a function of $U$ in
Fig. \ref{rp_rho} together with the corresponding value of  $\tilde U\tilde\rho(0)$.
The real and imaginary parts of the local dynamic spin susceptibility as calculated from the RPT formula
 are shown in Fig.\ref{compchis} for $U=5.6$ with the corresponding directly calculated NRG-DMFT results.
The NRG-DMFT results are not exact due to errors due to discretization and the broadening that has
to be introduced to give a continuous curve. The results can be seen to be in very reasonable
agreement.\par
\bigskip
\bigskip
\begin{figure}[!htbp]
\begin{center}
\includegraphics[width=0.38\textwidth]{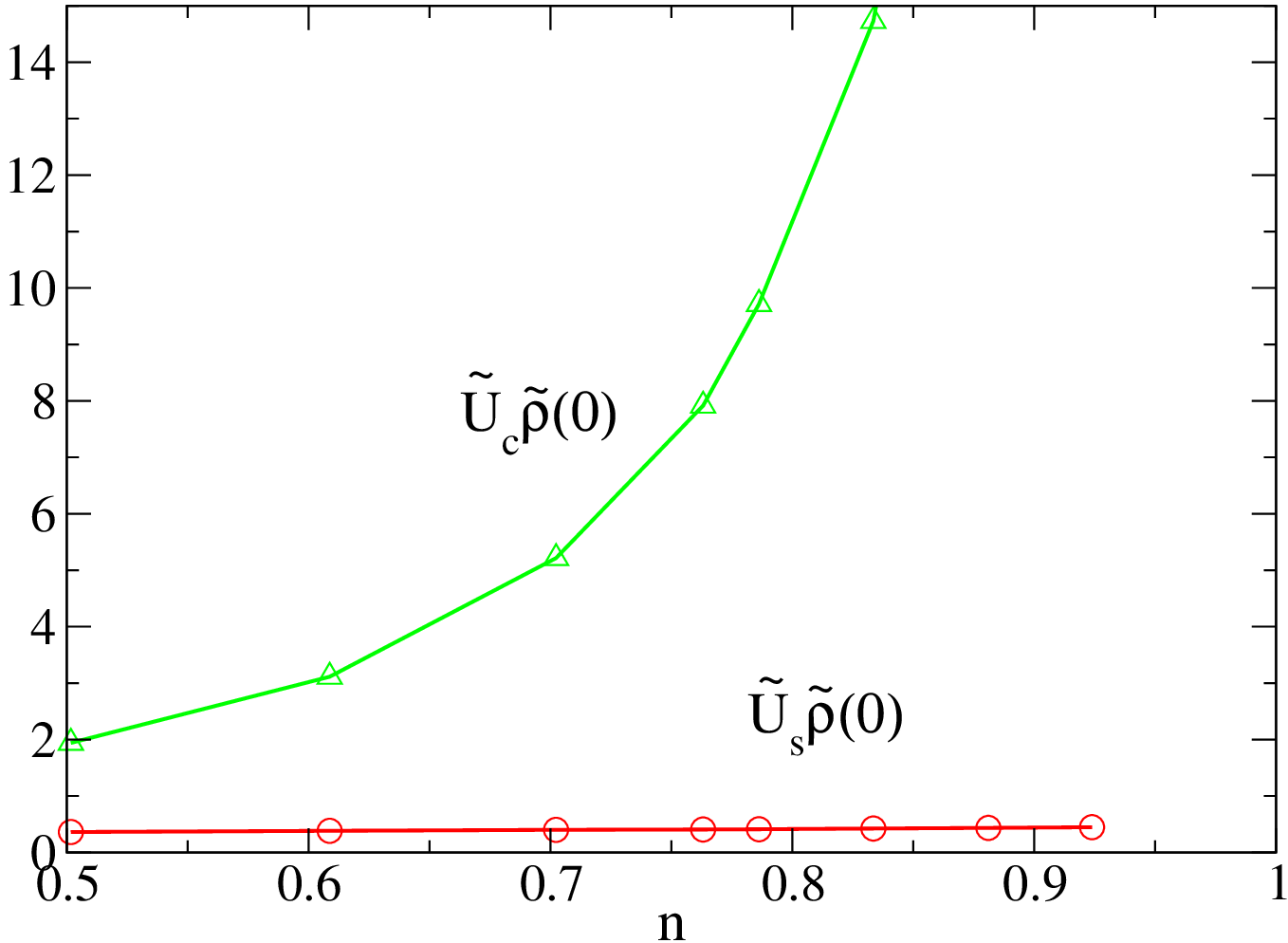}
\caption{
 Plots of $\tilde U_s\tilde\rho(0)$ and $\tilde U_c\tilde\rho(0)$
as a function of the filling factor $n$ for $U=6.0$.
 }
\vspace*{1.0cm}
\label{asym3}
\end{center}
\end{figure}
\noindent

\begin{figure}[!htbp]
\begin{center}
\includegraphics[width=0.4\textwidth]{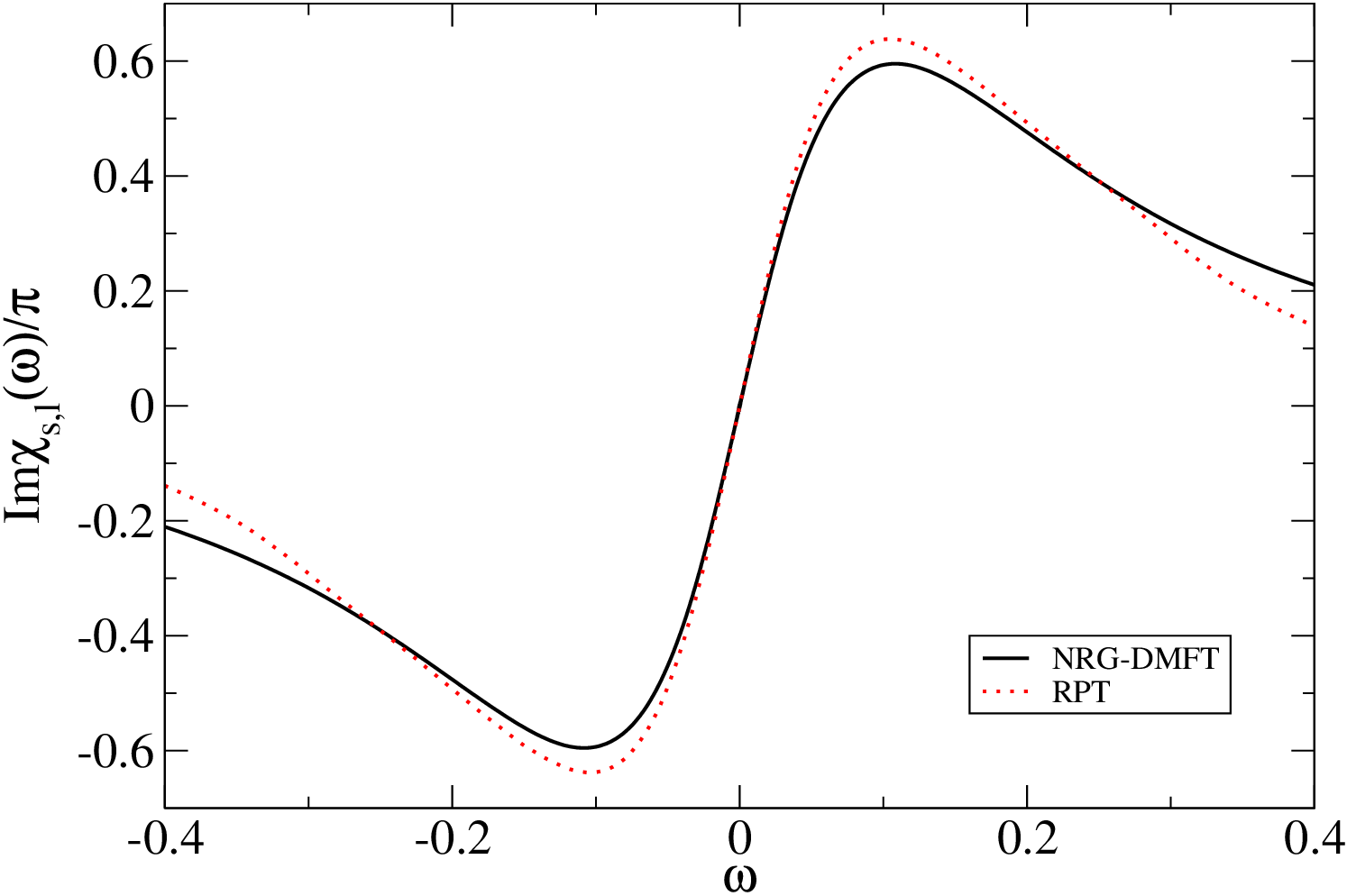}
\vspace*{0.5cm}
\caption{
A comparison of imaginary part of the RPT local dynamical
spin susceptibility  for $U=5.0$, $n=0.942$, with the corresponding NRG-DMFT results
 }
\label{ichiw_0.95_5.0}
\vspace*{1.0cm}
\end{center}
\end{figure}
\noindent

\begin{figure}[!htbp]
\begin{center}
\includegraphics[width=0.45\textwidth]{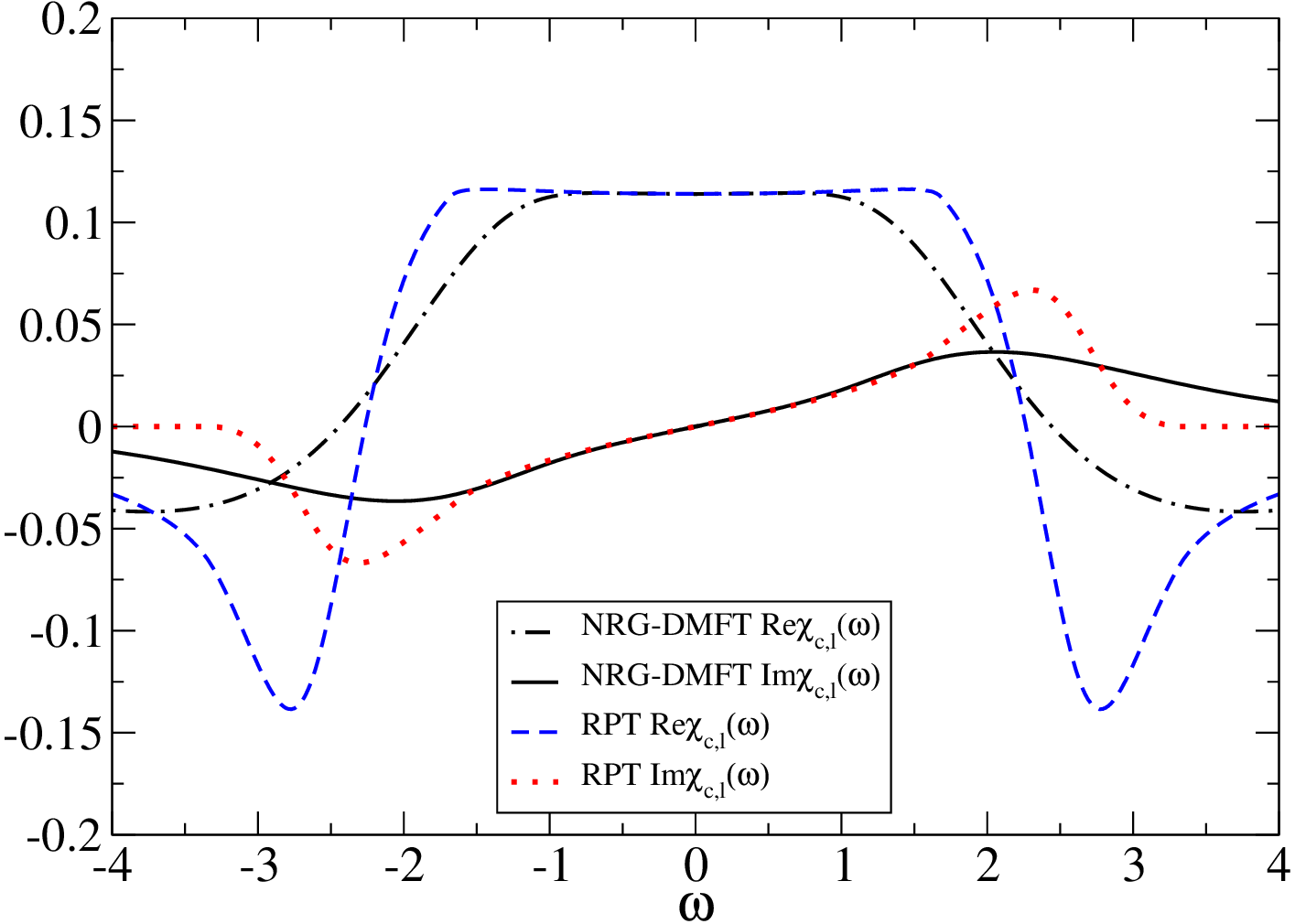}
\vspace*{0.5cm}
\caption{
A comparison of imaginary part of the RPT local dynamical
charge susceptibility  for U=1.5 at half-filling with the corresponding NRG-DMFT results
 }
\label{compc}
\vspace*{1.0cm}
\end{center}
\end{figure}
\noindent

In Fig. \ref{asym3}  the values of  $\tilde U_s\tilde\rho(0)$ and $\tilde U_c\tilde\rho(0)$
are shown away from half-filling as a function of the electron density $n$ for $U=6.0$.
The increase of $\tilde U_c\tilde\rho(0)$ as the density increases  reflects  the
lack of phase space for charge fluctuations when $U$ is close to or greater than $U_c$.
The RPT and NRG-DMFT results for the imaginary part of the local dynamic spin susceptibility
for the case  $U=5.0$, $n=0.942$
 are shown in Fig. \ref{ichiw_0.95_5.0}, and seen to be in good agreement. 
As the charge susceptibility is heavily suppressed for large value of $U$, NRG-DMFT and RPT results for 
the real and imaginary parts of the dynamic local charge susceptibility have been calculated  for a smaller value of $U$,  $U=1.5$, and are compared in Fig. \ref{compc}. Again over the low energy range there is general agreement
in the two sets of results.

\section{Calculation of $\chi_s({\bf q},\omega)$ and  $\chi_c({\bf q},\omega) \label{s6}$
}

 Here we discuss briefly the possibility of calculating the ${\bf q}$ and $\omega$ dependent
susceptibilities given information about the renormalized quasiparticles.
For the previous calculations it was sufficient to know only the local density of states $D(\omega)$
for the lattice and we used the form corresponding to a Bethe lattice. However for the calculation of the  $({\bf q},\omega)$ dependent susceptibilities one needs
the details of dispersion of the Bloch states $\epsilon_{\bf k}$. For this type of calculation the Bethe lattice,
 and even the hypercubic lattice for $d=\infty$, are inappropriate due to their special and restricted ${\bf k}$ dependence (for a discussion of this in detail see the review article of  Georges et al. \cite{GKKR96}). However, the DMFT is used as an approximation for calculations in the strong correlation regime
for a Hubbard model in three dimensions, and we could consider, for example, an $\epsilon_{\bf k}$ for a tight-binding cubic lattice. A much used approach for calculating the $({\bf q},\omega)$ spin susceptibility $\chi_s({\bf q},\omega)$
is the random phase approximation (RPA), which takes the form,
\begin{equation}
\chi_s({\bf q},\omega)={\chi^0({\bf q},\omega)
\over 1 - U\chi^0({\bf q},\omega)},
\label{rpas}
\end{equation}
where 
\begin{equation}
\chi^0({\bf q},\omega)= \sum_{\bf k} {f(\epsilon({\bf k}+{\bf
  q}))-f(\epsilon({\bf k}))\over
  (\omega-\epsilon({\bf k}+{\bf q})+\epsilon({\bf k}))}.
\label{rpapi}
\end{equation}
is the dynamic susceptibility of the free electrons.
The RPA approximation has been used recently, for example, to estimate the effective electron interaction due to spin fluctuations
\cite{HCW14}. This calculation is based on a perturbation expansion in powers of the bare interaction $U$.
It is of interest to see how this formula would be modified if the renormalization of the local interaction and of the
quasiparticles is taken into account. In the calculation of the dynamical susceptibilities for impurity
problems these renormalization effects are found to be very signficant\cite{Hew06}. They can be taken into account by
replacing $U$ by $\tilde U_s$, the renormalized interaction in the spin channel, and the replacing the dynamic susceptibility of the free electrons by the
corresponding susceptibility of the free quasiparticles, to give
\begin{equation}
\chi_s({\bf q},\omega)={\tilde\chi^0({\bf q},\omega)
\over 1 -\tilde U_s\tilde\chi^0({\bf q},\omega)},
\label{rpts}
\end{equation}
where 
\begin{equation}
\tilde\chi^0({\bf q},\omega)= \sum_{\bf k} {f(\tilde\epsilon({\bf k}+{\bf
  q}))-f(\tilde\epsilon({\bf k}))\over
  (\omega-\tilde\epsilon({\bf k}+{\bf q})+\tilde\epsilon({\bf k}))}.
\label{rptpi}
\end{equation}
The renormalized interaction $\tilde U_s$ is not simply $\tilde U$, as the series
of diagrams for $\omega=0$ contribute to the 4-vertex at zero frequency, and must
be cancelled by the counter term $\lambda_3$ so   $\tilde U_s=\tilde U-\lambda_3$.
The counter term $\lambda_3$ can be deduced from the calculated static uniform susceptibility
$\chi_s$ in Eqn. (\ref{chis}) as $\chi_s={\rm lim}_{{\bf q}\to 0}{\rm lim}_{{\omega}\to 0}\chi_s({\bf q})$.
As in the RPA this approximation assumes a local scattering vertex and 
 goes over to the the RPA result in the weak
correlation limit as $z\to 1$ and $\tilde U_{s}/z\to U$.
 With this formula, however, we can get enhanced low
energy spin fluctations for $\tilde U_{s}>0$ arising either close to the onset of a ferromagnetic
instability, which requires $\tilde U_{s}\tilde\rho(0)\ge 1$, where $\tilde\rho(0)$ is the
value of the quasiparticle density of states at the Fermi level, or close
to localization such that $z\ll 1$.  As in the RPA, in the case of a tight-binding cubic lattice at half
filling, an antiferromagnetic instability is predicted for   $\tilde U_{s}>0$ and s-wave superconductivity 
for  $\tilde U_{s}<0$.
\par
The charge susceptibility  $\chi_c({\bf q},\omega)$ can be calculated 
in a similar way, 
\begin{equation}
\chi_c({\bf q},\omega)={\tilde\chi^0({\bf q},\omega)
\over 1 +\tilde U_c\tilde\chi^0({\bf q},\omega)},
\label{rptc}
\end{equation}
Note, however, that unlike the standard RPA,  the interaction vertex $\tilde U_c$ is not in general the same as that in the spin channel. A similar approach, with RPA-like forms, with different vertices in the spin and charge
channels has been applied by Vilk and Tremblay\cite{VT97} to the two-dimensional Hubbard model 
to interpret the results of a Monte Carlo calculation. 
\par

\section{Summary \label{s7}} 

We have shown how information about the low energy quasiparticles can be deduced from an
analysis of the low energy fixed point in a DMFT calculation for the Hubbard model, and in
particular the on-site renormalized quasiparticle interaction $\tilde U$. This information
is sufficient to set up a renormalized perturbation expansion for the local self-energy
$\Sigma(\omega)$, which is applicable in all parameter regimes. It is particularly useful
to be able to derive analytic results in the very strong correlation limit where it is
difficult to obtain accurate results
from discrete sets of numerical data for the low energy spectra, or where some form of
broadening has been applied. We have been able to check some of the analytic expressions
in different regimes against the numerical results. We conjecture that there are some universal
relations on the approach to the Mott-Hubbard transition such that all the parameters
can be expresssed in terms of a single energy scale $T^*$ where $T^*\to 0$ at the transition.
 \par
The calculation of the renormalized parameters has been based on the assumption that the low
energy fixed point corresponds to a Fermi liquid. This appears to be the case in all the
regimes considered but the quasiparticles disappear on the approach to the Mott-Hubbard transition,
so the Fermi liquid expressions are only expected to be valid for temperatures $T$ such that
$T\ll T^*$. This leaves open the possibility of non-Fermi liquid behavior in the vicinity
of the Mott-Hubbard transition, as a quantum critical point, for temperatures such that
$T>T^*$. \par

The DMFT approach, with an on-site renormalized vertex $\tilde U$, is sufficient to carry out a renormalized
perturbation expansion for the self-energy of the infinite dimensional model. The characteristic feature of strongly correlated
electron systems is the strong
frequency dependence of the self-energy which is taken into account in the DMFT but at the expense of neglecting  any wavevector 
dependence. This  is a good initial approximation, taking into account the larger energy scale effects of strong electron correlation,  but in three and, particularly two dimensions, 
the wavevector dependence should be taken into account to examine  the  more subtle correlation effects that take place on the lowest energy scales.  An approach along related lines to that presented here  is
the dynamical vertex approximation\cite{TKH07} (D$\Gamma$A), which involves estimates of both the
frequency and wavevector dependence of the irreducible 4-vertices. A recent application of this approximation to the Hubbard model is that of Rohringer and Toschi\cite{RT16}. A simplified feature of the RPT calculation of the
low energy response functions on the lowest energy scales is the neglect of the frequency
dependence of these renormalized vertices. This gives excellent results, for example,  in the strong correlation regime for the Anderson impurity model\cite{Hew06}. Some estimate  
of the ${\bf q},\omega$ dependent
spin susceptibility, based on a generalized  RPA with a local scattering vertex and renormalized parameters derived from a DMFT-NRG calculation,  was
outlined in section \ref{s6}.
A reasonable approximation
going beyond the local approximation could be to take nearest neighbour contributions for the renormalized
four-vertex into account, and again neglect any frequency dependence. It is important, however,
in using any renormalized vertex that counterterms have to be taken into account to prevent over-counting.

\bigskip
\bigskip\noindent{\bf Acknowledgement}\par
\bigskip
\noindent
We wish to thank Winfried Koller, Dietrich Meyer and Johannes Bauer for their contributions to the
development of the NRG used in the calculations, and Sriram Shastry for helpful discussion and comments.

\bibliography{artikel}

\end{document}